\documentclass[a4paper]{llncs}
\usepackage{amssymb,latexsym,amsxtra,amscd,amsfonts}
\begin{document}
\pagestyle{plain}


\newcommand{\cO}{{\mathcal O}}
\newcommand{\cE}{{\mathcal E}}
\newcommand{\cI}{{\mathcal I}}
\newcommand{\cF}{{\mathcal F}}
\newcommand{\cR}{{\mathcal R}}

\newcommand{\Z}{\mathbb Z}
\newcommand{\N}{\mathcal N}
\newcommand{\cS}{\mathcal S}
\newcommand{\cL}{{\mathcal L}}
\newcommand{\F}{\mathbb F}
\newcommand{\C}{\mathbb C}
\newcommand{\D}{\mathbb D}
\newcommand{\Q}{\mathbb  Q}
\newcommand{\G}{\mathbb G}
\newcommand{\m}{\mathfrak m}
\newcommand{\n}{\mathfrak n}
\newcommand{\f}{\mathfrak f}
\newcommand{\e}{\mathfrak e}
\newcommand{\g}{\mathfrak g}
\newcommand{\h}{\mathfrak h}
\newcommand{\be}{\mathfrak b}

\newcommand{\q}{\mathfrak q}
\newcommand{\pe}{\mathfrak p}
\newcommand{\ma}{\mathfrak a}

\newcommand{\mF}{\mathcal F}
\newcommand{\mG}{\mathcal G}

\newcommand{\tF}{\tilde{\mathcal F}}
\newcommand{\tG}{\tilde{G}}
\newcommand{\tx}{\tilde{x}}
\newcommand{\td}{\tilde{d}}
\newcommand{\tf}{\tilde{f}}
\newcommand{\ta}{\tilde{a}}
\newcommand{\te}{\tilde{e}}
\newcommand{\tM}{\tilde{M}}
\newcommand{\txF}{\widetilde{x_iF}}
\newcommand{\tI}{\tilde{I}}

\newcommand{\Fp}{\F_p}
\newcommand{\Fq}{\F_q}

\newcommand{\ka}{{\kappa}}
\newcommand{\reg}{\rm reg}
\newcommand{\End}{\rm End}
\newcommand{\Pic}{\rm Pic}
\newcommand{\NS}{\rm NS}
\newcommand{\Div}{\rm div}
\newcommand{\nuinf}{\nu_{\infty}}
\newcommand{\siminf}{\sim_{\infty}}
\newcommand{\bu}{\bf u}
\newcommand{\bt}{\bf \theta}
\newcommand{\btt}{\bf t}
\newcommand{\bz}{\bf z}
\newcommand{\la}{\langle}
\newcommand{\ra}{\rangle}

\title{Weil descent and cryptographic trilinear maps}
\author{Ming-Deh A. Huang (USC, mdhuang@usc.edu)}
\institute{Computer Science Department,University of Southern California, U.S.A.}


\maketitle

\begin{abstract}
It has recently been shown that cryptographic trilinear maps are sufficient for achieving indistinguishability obfuscation.
In this paper we develop a method for constructing such maps
on the Weil descent (restriction) of abelian varieties over finite fields, including
the Jacobian varieties of hyperelliptic curves and elliptic curves.
The security of these candidate cryptographic trilinear maps raises several interesting questions, including the computational complexity of a trapdoor discrete logarithm problem.
\end{abstract}
\section{Introduction}
Cryptographic applications of multilinear maps beyond bilinear maps were first proposed in the work of  Boneh and Silverberg \cite{BS}.  However the existence of cryptographically interesting $n$-multilinear maps for $n > 2$ remains an open problem. The problem has attracted much attention more recently as multilinear maps and their variants have become a useful tool for indistinguishability obfuscation.
Initially coined in the work of Barak et al. \cite{BGI}, indistinguishability obfuscation is a powerful notion with sweeping applications and far reaching consequences in cryptography.
Very recently Lin and Tessaro \cite{LT} showed that trilinear maps are sufficient for the purpose of achieving indistinguishability obfuscation (see \cite{LT} for references to related works along several lines of investigation).
The striking result of Lin and Tessaro \cite{LT} has brought the following question into the spotlight: can a cryptographically interesting algebraic trilinear map be constructed?
In this paper we develop a method for constructing such trilinear maps
on the Weil descent (restriction) of abelian varieties over finite fields, including
the Jacobian varieties of hyperelliptic curves and elliptic curves.

A natural place to look for $n$-multilinear maps for $n >2$ is \'{e}tale cohomology.  The challenge however is identifying a promising candidate in the abstract form of \'{e}tale cohomology that may have concrete and efficient realization.   For example, Huang and Raskind \cite{HR}
gave $n$-multilinear generalization of Tate pairing under suitable conditions, for $n$ even.  However the generalized $n$-multilinear Tate pairing naturally takes values in $n/2$-fold tensor product of $\mu_{\ell}$, the group of $\ell$-th roots of unity, therefore requires solving CDH (computational Diffie-Hellman) problem when $n > 2$  to consolidate the value of pairing in $\mu_{\ell}$.  In fact Boneh and Silverberg (see Corollary 7.6 in \cite{BS}) gave necessary conditions that seem difficult to satisfy for Galois-equivariant $n$-multilinear maps taking values in $\mu_{\ell}$.
However Chinburg (at the AIM workshop on cryptographic multilinear maps (2017)) recently demonstrated a trilinear map taking values in $\mu_{\ell}$ can be derived from \'{e}tale cohomology, and this was the starting point of our line of investigation.

Following up on Chinburg's observation we take the following approach to construct trilinear maps.
We start with a principally polarized  abelian variety over a finite field and make use of the pairing of the torsion points, as well as the action of endomorphisms on the torsion points to construct a trilinear map.  To strengthen the security of the third pairing group, which acts on the second pairing group through endomorphisms,
we apply the idea of Weil descent (or Weil restriction) \cite{A,FL}.
The security of the trilinear maps constructed in this paper raises several interesting questions, including the computational complexity of a trapdoor discrete logarithm problem.

\subsection{General idea of construction}
Our line of investigation was motivated by an observation of Chinburg that the following map from \'{e}tale cohomology may serve as the basis of constructing a cryptographically interesting trilinear map:

\[ H^1 (A,\mu_{\ell})\times H^1 (A,\mu_{\ell})\times H^2 (A,\mu_{\ell})\to H^4 (A, \mu_{\ell}^\otimes{3})\cong \mu_{\ell}\]
where $A$ is an abelian surface over a finite field $\F$ and the prime $\ell\neq {\rm char}(\F)$.  This trilinear map is the starting point of the following more concrete construction.

Suppose $A$ is a principally polarized abelian variety over a finite field $\F$. Let $A^*$ denote the dual abelian variety. Consider $A$ as a variety over $\bar{\F}$, the algebraic closure of $\F$.
Let $e_{\ell}$ be the pairing between $A[\ell]$ and $A^* [\ell]$ (\cite{MilneA} \S~16).

In \cite{H} the following trilinear map $(\alpha,\beta,\cL) \to e_{\ell}(\alpha,\varphi_{\cL} (\beta))$ was considered, where  $\alpha,\beta\in A[\ell]$, $\cL$ is an invertible sheaf, and $\varphi_{\cL}$ be the map $A\to A^* =\Pic^0 (A)$ so that
\[ \varphi_{\cL} (a) =t_a^*\cL \otimes \cL^{-1} \in \Pic^0 (A)\]
for $a\in A(\bar{\F})$ where $t_a$ is the translation map defined by by $a$ (\cite{MilneA} \S~1 and \S~6).

Note that in the map just described we no longer need to assume that $A$ is of dimension 2.

Below we describe the general idea of constructing a cryptographic trilinear map motivated by the above discussion.

Our goal is to construct an $\F_{\ell}$-linear map $G_1\times G_2\times G_3\to G_4$ with $G_i\cong \Z/\ell\Z$ as groups for $i=1,\dots,4$.  The basic requirement is that the discrete logarithm problems on the four groups are computationally hard while the trilinear map is efficient to compute.

The basic setup of our construction can be described as follows.  Suppose $V$ is a finite dimensional vector space over $\F_{\ell}$ with an efficiently computable pairing $e:V\times V \to \mu_{\ell}$. Let $\End  V$ denote the ring of endomorphism of $V$ as an $\F_{\ell}$-vector space. We choose $\alpha,\beta\in V$ with $e(\alpha,\beta)\neq 1$, and set $G_i$ as the group generated by $\alpha$ and $\beta$ respectively for $i=1$ and $2$.   Let $E_0 = \{ \lambda\in\End V: e(\alpha,\lambda\beta)=1\}$.   We form the third group $G_3$ as a quotient $U_1/U$ where $U$ is a submodule of $E_0$ and $U_1 = \F_{\ell}+ U$.  Then we have a trilinear map $G_1\times G_2\times G_3\to \mu_{\ell}$ sending $(a\alpha,b\beta, c+\lambda)$ to
$e(a\alpha,c+\lambda(b\beta)) = e(\alpha,\beta)^{abc}$ for $a,b,c\in\F_{\ell}$ and $\lambda\in U$.

In our construction $V=A[\ell]$ the set of $\ell$-torsion points of an abelian variety $A$ over a finite field.  We assume $A$ is simple and  principally polarized.
Let $e: A[\ell]\times A[\ell]\to\mu_{\ell}$ be a non-degenerate skew-symmetric pairing.  An important example is the pairing defined by a principal polarization of $A$ and the canonical pairing between $\ell$-power torsion points of $A$ and the dual abelian variety.  We will need to make sure that the pairing $e$ is efficiently computable.   For now suppose this is the case.  We find $\alpha,\beta\in A[\ell]$ such that
$e(\alpha,\beta) \neq 1$, and let $G_1$ and $G_2$ be respectively the cyclic groups generated by $\alpha$ and $\beta$.

As the first attempt we may take $G_3$ as a quotient $W_1/W$ where $W_1 =\F_{\ell}+W$ and $W$ is a submodule of
$E_0=\{ \lambda\in\End (A[\ell]):
e (\alpha,\lambda(\beta) )=1\}$.
And we have a trilinear map $G_1\times G_2 \times G_3 \to \mu_{\ell}$ sending
$(a\alpha,b\beta,c+W)$ to $\zeta^{abc}$ where $\zeta=e(\alpha,\beta)$.

We need a representative $f\in c+W$ to be specified in such a way that $f$ can be efficiently applied to points in $G_2$. On the other hand given $f$ that represents an element of $U_1$, we want it to be hard to determine $c\in\F_{\ell}$ such that $f\in c+U$.  This can be a problem given that the $\F_{\ell}$-dimension of $W_1$ is bounded by that of $\End (A[\ell])$, which is $O(g^2)$ where $g=\dim A$.   More generally there can be a problem if $G_3$ is presented explicitly as a quotient $U_1/U$ where $U_1$ is a subspace of some $\F_{\ell}$ vector space of polynomially bounded dimension.
The reason is that
in cryptographic applications we often need to assume that polynomially many samples from $U$ are revealed to the public (hence the adversary).   If the dimension of $U$ is polynomially bounded then  a basis of $U$ can likely be determined from the sampled elements of $U$.
The basis of $U$ together with 1 form a basis of $U_1$.  Now the problem of finding $c$ such that $f-c\in U$ is easy.

Our strategy in meeting the challenge involves Weil descent (or Weil restriction) \cite{A,FL}. Weil descent was introduced by Frey \cite{F} as a constructive tool in cryptography to disguise elliptic curves.  In \cite{DG}
Dent and Galbraith applied the idea to construct trapdoor DDH (Decision Diffie-Hellman) groups by disguising elliptic curves in order to hide pairings.
In our approach, Weil descent is involved for deeper reasons than disguising elliptic curves or abelian varieties, as will be seen in our discussions below.
We remark that the construction in \cite{DG} is vulnerable to the attacks described in \cite{Mo}, which depend critically on the addition morphism of an elliptic curve of interest being given in the projective model by homogeneous polynomials.
The attacks do not extend to our constructions, where the abelian varieties and maps are given strictly by affine models in affine pieces.

We now give a brief outline of our approach.

We start with an abelian variety defined over a extension $K$ of finite field $k$ with $d=[K:k]$,  however we proceed to construct the trilinear map on a Weil descent $\hat{A}$ of $A$.  The Weil descent $\hat{A}$ is formed with respect to a secret basis of $K$ over $k$.  Now $\hat{A}[\ell]$ is isomorphic to $A[\ell]^d$, so $\End \hat{A}[\ell]$ contains a submodule isomorphic to $Mat_d (\F_{\ell})$, the algebra of $d$ by $d$ matrices over $\F_{\ell}$.    Utilizing the secret basis we select
a set $S$ of $N=d^{O(1)}$ elements $\lambda_i\in \End \hat{A}[\ell]$ such that $\lambda_i$ corresponds to a matrix $M_i \in Mat_d (\F_{\ell})$.  We consider $\hat{A}[\ell]$ a blinded version of $A[\ell]^d$, and $\lambda_i$ a blinded version of $M_i$.  The maps $\lambda_i$ will be specified in such a way that they can be efficiently applied to $\hat{A}[\ell]$ while the matrices $M_i$ are hidden.   Our trilinear map will be derived from a blinded version of the following trilinear map:
\[
\begin{array}{rcl}
A[\ell]^d \times A[\ell]^d \times Mat_d (\F_{\ell}) & \to & \mu_{\ell}\\
(\alpha,\beta,M) &\to & e(\alpha, M(\beta))
\end{array}
\]
where $\alpha,\beta\in A[\ell]^d$, $M\in Mat_d(\F_{\ell})\subset \End (A[\ell]^d)$, and $e$ is a non-degenerate bilinear pairing on $A[\ell]^d$ (determined by a non-degenerate bilinear pairing on $A[\ell]$).

Let $\Lambda$ be the $\F_{\ell}$-algebra generated by $N$ independent variables $z_1,\ldots,z_N$, which is non-commutative for $N > 1$.  Let $\Lambda$ act on $\hat{A}[\ell]$ such that $z_i$ acts as $\lambda_i$ for $i=1,\ldots,N$.
We have a morphism of algebras $\lambda: \Lambda\to Mat_d (\F_{\ell})$ such that $\lambda(z_i)=M_i$, for $i=1,\ldots,N$, serving as a secret trapdoor map.  The following trilinear map can be considered a blinded version of the trilinear map just described:

\[
\begin{array}{rcl}
\hat{A}[\ell] \times \hat{A}[\ell] \times \Lambda & \to & \mu_{\ell}\\
(\hat{\alpha},\hat{\beta},f) &\to & \hat{e}(\hat{\alpha}, f(\lambda_1,\ldots,\lambda_N)(\hat{\beta}))
\end{array}
\]
where $\hat{\alpha},\hat{\beta}\in\hat{A}[\ell]$, $f\in\Lambda$, and $\hat{e}$ is pairing on $\hat{A}[\ell]$ which is a blinded version of $e$.

To construct our trilinear map we take $G_1$ (resp. $G_2$) to be the cyclic group generated by a point $\hat{\alpha}$ in $\hat{A}[\ell]$ (resp. $\hat{\beta}\in\hat{A}[\ell]$).   To construct the third pairing group $G_3$,
we form a set $R_1$ of relations of degree 2 on $M_i$ (hence $\lambda_i$), and publish the set $\cR$ of relations on $z_i$ whose image under $\lambda$ is $R_1$.
Let $J$ be the two-sided ideal of $\Lambda$ generated by $\cR$, and let
$J_N$ be the submodule of $J$ consisting of elements of degree less than or equal to $N$.
With the action of $Mat_d (\F_{\ell})$ on $A[\ell]^d$ blinded by the action of $\Lambda$ on $\hat{A}[\ell]$, we define the third pairing group $G_3$ in terms of modules in $\Lambda$ of exponential dimensions over $\F_{\ell}$.
We set $G_3=U_1/U\cong \Z/\ell\Z$ where $U=J_N$ and $U_1 = \F_{\ell}+U$.
For $a\in\F_{\ell}$, $a+U\in G$ is encoded by a sparse representative $g$ in $a+U$.  Given $g$ to determine $a$ seems hard as the dimension of $U$ is exponentially large.  Using $\cR$ one can formulate a system of quadratic polynomials in $d^{O(1)}$ variables to determine $\lambda$, however solving such a system is too costly.  If the secret descent basis is uncovered, then the trapdoor map $\lambda$ can be efficiently determined, and the discrete logarithm problem on $G_3$ is reduced via $\lambda$ to $Mat_d (\F_{\ell})$, which is a vector space of polynomially bounded dimension.  Therefore the security of the trilinear map depends on the descent basis being a trapdoor secret, and a methodology is needed to specify maps and functions on a descent variety while protecting the secrecy of the descent basis.  This is the subject of investigation in the next section.

\section{Weil descent and secrecy}
Let $k$ be a finite field of $q$ elements and let $K$ be an extension of degree $d$ over $k$.
Let $\bt$ denote a public basis of $K$ over $k$ consisting of
$\theta_0,\ldots,\theta_{d-1}\in K$.  Every published element of $K$ is explicitly written in this basis.
Let $\bu$ denote a secret basis of $K$ over $k$ consisting of
$u_0,\ldots,u_{d-1}\in K$.  The basis $\bu$ is the basis with respect to which descent objects are defined.
The Galois group $G(K/k)$ is generated by the Frobenius automorphism $\sigma$ such that for $x\in K$, $\sigma(x)=x^q$.
For $i\in\Z$, let $\sigma_i =\sigma^{i\mod d}$.

For $\hat{x},\hat{y}\in\bar{k}^d$, let $\la \hat{x},\hat{y}\ra = \sum_{i=0}^{d-1} x_i y_i$ where $\hat{x}=(x_i)_{i=0}^{d-1}$ and
$\hat{y}=(y_1)_{i=0}^{d-1}$.

Let $\delta:\bar{k}^d\to\bar{k}$ such that for $\hat{x}=(x_j)_{j=0}^{d-1}$ with $x_j \in \bar{k}$,
$\delta (\hat{x}) = \sum_{j=0}^{d-1} x_j u_j = \la \hat{x}, \bu\ra.$

We have $\delta^{\sigma_i} (\hat{x})= \sum_{j=0}^{d-1} x_j u_j^{\sigma_i} = \la \hat{x}, \bu^{\sigma_i}\ra.$

Let $\rho: \bar{k}^d\to\bar{k}^d$ such that for $\hat{x}\in\bar{k}^d$, $\rho (\hat{x}) = (\delta^{\sigma_i} (\hat{x})_{i=0}^{d-1}$.

Let $\Gamma = (u^{\sigma_i}_j)_{0\le i,j\le d-1}$.
Let $W=\Gamma^{-1}=(w_{ij})_{0\le i,j\le d-1}$.

Throughout this section it will be useful to consider the basis $\bu$ as being secret, hence the maps $\delta$ and $\rho$ and the matrices $\Gamma$ and $W$ are secret as well.

A point $\hat{x}\in\bar{k}^d$ is called a {\em descent point} if there is some $y\in\bar{k}$ such that
$\rho\hat{x}=(y^{\sigma_i})_{i=0}^{d-1}$.

\begin{lemma}
\label{descent-point}
A descent point that is not $k$-rational reveals information about $\bu$ in the sense that
$\la \hat{x} - \hat{x}^{\sigma_{i}} , \bu\ra = 0$ for all $i$.
\end{lemma}
\ \\{\bf Proof}  Suppose $y\in\bar{k}$ is such that $\rho\hat{x} = (y^{\sigma_i})_{i=0}^{d-1}$.  Since
$y=\la \hat{x}, \bu\ra$, we get $y^{\sigma_i} = \la \hat{x}^{\sigma_i}, \bu^{\sigma_i}\ra$.  On the other hand we also have $y^{\sigma_i} = \la \hat{x},\bu^{\sigma_i} \ra$ since $\rho\hat{x} = (y^{\sigma_i})_{i=0}^{d-1}$.  Therefore
$\la \hat{x} - \hat{x}^{\sigma_{i}} , \bu\ra = 0$. $\Box$

We consider descent points weak in light of Lemma~\ref{descent-point}.

Suppose $F\in K[x_1,\ldots,x_n]$.  Let $\hat{x}_i = (x_{ij})_{j=0}^{d-1}$, for $i=1,\ldots,n$.
Then
\[ F (\delta (\hat{x}_1),\ldots,\delta (\hat{x}_n) ) = \sum_{i=0}^{d-1} f_i (\hat{x}_1,\ldots,\hat{x}_n) u_i ,\]
where $f_i \in k [\hat{x}_1,\ldots,\hat{x}_n]$.  We denote by $\hat{F}$ the tuple $(f_i)_{i=0}^{d-1}$.

If we identify $\bar{k}^{dn}$ as the $n$-fold product $\bar{k}^d \times \ldots \times \bar{k}^d$, and by abuse of notation denote $\delta$ as the map $\bar{k}^{dn} \to \bar{k}^n$ such that
$\delta (\hat{x_1},\ldots,\hat{x}_n) = (\delta (\hat{x}_1),\ldots, \delta(\hat{x}_n))$ where $\hat{x}_1,\ldots,\hat{x}_n\in\bar{k}^d$.

Put $\hat{X}=\hat{x_1},\ldots,\hat{x}_n$.  Let $\hat{F} = (f_i)_{i=0}^{d-1}$.  Then we may write
\[ F(\delta(\hat{X})) = \delta (\hat{F}(\hat{X})) = \la \hat{F} (\hat{X}), \bu\ra .\]
We have
\[ F^{\sigma_i}(\delta^{\sigma_i}(\hat{X})) = \delta^{\sigma_i} (\hat{F}(\hat{X})) = \la \hat{F} (\hat{X}), \bu^{\sigma_i}\ra .\]

Consider the map $F:\bar{k}^n \to \bar{k}$ defined by $F$ sending $(x_1,\ldots,x_n)\in \bar{k}^n$ to  $F(x_1,\ldots,x_n)$. Then $\hat{F}$ defines a map $\hat{F}:\bar{k}^{nd}\to\bar{k}^d$ sending
$(\hat{X}) = (\hat{x}_1,\ldots,\hat{x}_n)\in\bar{k}^{nd}$, with $\hat{x}_i\in\bar{k}^d$, to $\hat{F} (\hat{X})$.

We have the following commutative diagrams
\[
\begin{array}{ccc}
\begin{array}{lll}
\bar{k}^{nd} & \stackrel{\hat{F}}{\to} &\bar{k}^d\\
\downarrow \delta^{\sigma_i} &   & \downarrow \delta^{\sigma_i}\\
\bar{k}^n & \stackrel{F^{\sigma_i}}{\to} & \bar{k}

\end{array}

&

&

\begin{array}{lcl}
\bar{k}^{nd} & \stackrel{\hat{F}}{\longrightarrow} &\bar{k}^d\\
\downarrow \rho &   & \downarrow \rho\\
\bar{k}^{nd} & \stackrel{\prod_{i=0}^{d-1} F^{\sigma_i}}{\longrightarrow} & \bar{k}^{d}

\end{array}

\end{array}
\]

In particular $\hat{F}=\rho^{-1}\circ\prod_{i=0}^{d-1} F^{\sigma_i}\circ\rho$.

Let $R=K[x_1,\ldots,x_n]$ and $\hat{R}=k[\hat{x}_1,\ldots,\hat{x}_n]$.
Suppose $V=Z(F_1,\ldots,F_m)$, the algebraic set defined by the zeroes of $F_1,\ldots,F_m \in R$.
The {\em descent} $\hat{V}$ of $V$ with respect to $\bu$ is defined by $\hat{V} = Z (\hat{F}_1,\ldots,\hat{F}_m)$.

The map $\rho$ induces an isomorphism $\hat{V}\to\prod_{i=0}^{d-1} V^{\sigma_i}$.  Since $\bu$ is secret, $\rho$ is secret as well, so we may consider $\hat{V}$ as a blinded version of $\prod_{i=0}^{d-1} V^{\sigma_i}$, and consider a rational function $\phi:\hat{V}\to \bar{k}$ as a blinded version of
$\phi'=\phi\circ\rho^{-1}:\prod_{i=0}^{d-1} V^{\sigma_i}\to\bar{k}$.  Suppose $\alpha\in\hat{V}$ and $\rho (\alpha) = (\beta_i)_{i=0}^{d-1}$ with $\beta_i\in V^{\sigma_i}$.  Then $\alpha$ is the blinded point of $(\beta_i)_{i=0}^{d-1}$,
and $(\alpha, \phi(\alpha))$ is a blinded version of $(\beta,\phi'(\beta))$ where $\beta=(\beta_i)_{i=0}^{d-1}$.

Suppose $\varphi: V\to\bar{k}$ is a rational function defined over $K$.  We denote by $\hat{\varphi}$ the map
$\hat{\varphi}=\rho^{-1}\circ\prod_{i=0}^{d-1}\varphi^{\sigma_i}\circ\rho$, and we have the commutative diagram
\[\begin{array}{lcl}
\hat{V} & \stackrel{\hat{\varphi}}{\longrightarrow} &\bar{k}^d\\
\downarrow \rho &   & \downarrow \rho\\
\prod_{i=0}^{d-1} V^{\sigma_i} & \stackrel{\prod_{i=0}^{d-1} \varphi^{\sigma_i}}{\longrightarrow} & \bar{k}^{d}

\end{array}\]
We consider the map $\hat{\varphi}$ as a blinded version of $\prod_{i=0}^{d-1}\varphi^{\sigma_i}$.  Moreover we note that
$\varphi^{\sigma_i}\circ\delta^{\sigma_i}=\delta^{\sigma_i}\circ\hat{\varphi}$ is a blinded version of
$pr_i\circ\prod_{j=0}^{d-1}\varphi^{\sigma_j}=\varphi^{\sigma_i}\circ pr_i$ in the sense described above, where $pr_i$ denotes the projection to the $i$-th coordinate.

For the blinding to be effective we want to maintain the secrecy of $\bu$ while we specify $\hat{V}$, and $\hat{\varphi}$ or $\varphi^{\sigma_i}\circ\delta^{\sigma_i}$.

A few observations are in order.
\begin{enumerate}
\item For $\alpha\in\hat{V}$, if both $\alpha$ and $\delta\alpha$ are made public, then a linear relation on the $u_i$ is revealed:  $\la \alpha, \bu \ra = \delta\alpha$.
\item
In our setting we assume polynomially many points from $\hat{V}$ can be sampled.
If both $\hat{F}$ and $F\circ\delta$ are specified then for sampled $\alpha\in\hat{V}$, let $\beta = \hat{F} (\alpha)$, then $\delta\beta = \delta\hat{F} (\alpha) = F\circ \delta( \alpha)$, hence a a linear relation on $u_i$ is revealed from $\beta$ and $\delta\beta$.
\item If $F$ is known and $F\circ\delta$ is specified, then for a sampled $\alpha\in\hat{V}$, $F\circ\delta (\alpha)$ yields linear relation for a set of monomials in $u_0,\ldots,u_{d-1}$.  Take for example $F=axy$, then $F\circ\delta (\hat{x},\hat{y}) = \sum_{i,j} a u_i u_j x_i y_j$.   If
    $F\circ\delta (\alpha) = b$,  then
    we get the relation  $b=\sum_{i,j} a \alpha_i \alpha_j u_i u_j$ on $u_i u_j$, where $\alpha = (\alpha_i)_{i=0}^{d-1}$.

\end{enumerate}
Therefore we do not specify both $\hat{\varphi}$ and $\varphi\circ\delta$ (or $\varphi^{\sigma_i}\circ\delta^{\sigma_i}$), and we do not specify $\varphi\circ\delta$ if  $\varphi$ is known to the public.

In the following subsections we investigate more fully what and how descent algebraic sets and maps can be specified so as to maintain the secrecy of the descent basis $\bu$.

\subsection{Global descent}
\label{gd}
We call $(f_i)_{i=0}^{d-1}$, with $f_i \in k [\hat{x}_1,\ldots,\hat{x}_n]$, a {\em global descent} (with respect to $\bu$) if there is $F\in K[x_1,\ldots,x_n]$ such that $\hat{F} = (f_i)_{i=0}^{d-1}$.
We call a polynomial $G\in K[\hat{x}_1,\ldots,\hat{x}_n]$, a {\em K-global descent} for $\bu^{\sigma_i}$ if there is $F\in K[x_1,\ldots,x_n]$ such that $G =\delta^{\sigma_i}\circ  \hat{F}=F^{\sigma_i}\circ\delta^{\sigma_i}$.

\begin{lemma}
\label{kgld}
Let $H=(f_i)_{i=0}^{d-1}$ with $f_i \in k [\hat{x}_1,\ldots,\hat{x}_n]$.  Then
$\la H, \bu^{\sigma_i}\ra$ is a $K$-global descent for $\bu^{\sigma_i}$ if and only if $H$ is a global descent.
\end{lemma}
\ \\{\bf Proof} If $H$ is a global descent then $H=\hat{F}$ for some $F\in K[x_1,\ldots,x_n]$.
Then $\la H, \bu^{\sigma_i} \ra = \la \hat{F},\bu^{\sigma_i}\ra = \delta^{\sigma_i}\circ\hat{F}$ is a $K$-global descent.

Conversely if $\la H, \bu^{\sigma_i}\ra$ is a $K$-global descent, then there is some $F\in K[x_1,\ldots,x_n]$ such that
$\la H, \bu^{\sigma_i}\ra=\delta^{\sigma_i}\circ  \hat{F}$.  Since $\delta^{\sigma_i}\circ  \hat{F}=\la \hat{F},\bu^{\sigma_i} \ra$, we have $\la H-\hat{F}, \bu^{\sigma_i}\ra=0$.  Since $H-\hat{F}$ is fixed by $\sigma$,  we have $\la H-\hat{F}, \bu^{\sigma_j}\ra=0$ for $j=0,\ldots,d-1$.  This implies $H=\hat{F}$, given that $\rho$ is invertible.  $\Box$

Global descents and $K$-global descents are objects that reveal the identity of $\bu$, as shown in Proposition~\ref{hatF} and Proposition~\ref{tildeF} below.  Consequently they should not be formed and made public if $\bu$ is to remain secret.

Proposition~\ref{global-descent} and Proposition~\ref{K-global-descent} show that whether or not a tuple of polynomials contains any global descent and whether a polynomial contains any $K$-global descent can be efficiently checked using the basis $\bu$.

Proposition~\ref{gld} characterizes $Gl_d(k)$-action on a global descent, and shows in particular that for a global descent $\hat{F}$, the fraction of $\Gamma\in Gl_d (k)$ such that $\Gamma\hat{F}$ contains any global descent is negligible.  These results will be applied in the next subsection to show how descent maps on nontrivial descent varieties can be properly specified so as to keep $\bu$ secret.

We note that implicit in our notation is the assumption that the association between the variables in $\hat{x}_i$ and $x_i$, for all $i$, is public information.

A {\em term} $T$ with coefficient $a$ is of the form $am$ where $a$ is a constant and $m$ is a monomial. Call a term $T$ {\em vital} if it is of degree greater than 1 or of the form $ax_i$ where $K=k(a)$.

The {\em support} of a polynomial is the set of monomials that appear in the polynomial with nonzero coefficient.

For $a\in K$, let $\Gamma_a = (\gamma_{ij})$ be the $d$ by $d$ matrix in $Gl_d (k)$ such that
$au_i = \sum_{j=1}^d \gamma_{ij} u_j$.  Note that the fraction of $\Gamma\in Gl_d (k)$ such that $\Gamma=\Gamma_a^t$ for some $a\in K$ is in roughly
$\frac{|k|^d}{|k|^{d^2}}$, which is negligible.

\begin{proposition}
\label{hatF}
Suppose $F\in K[x_1,\ldots,x_n]$ contains a vital term.  Then given $\hat{F}$ one can efficiently uncover the descent basis.
\end{proposition}
\ \\{\bf Proof}
Write $F$ as the sum of terms $F=\sum T_i$.  Then $\hat{F}=\sum_i \hat{T_i}$.   From $\hat{F}$ we can read off $\hat{T_i}$ easily since $\hat{T_i}$ have disjoint supports, each determined completely by the corresponding monomial in $T_i$.  So it is enough to consider the case where $F$ is a vital term $T$.

Suppose $F=T$ is a vital term and for simplicity suppose  $T=a x_1\ldots x_r$ for some $r\ge 1$, where either $r > 1$ or $r=1$ and $K = k(a)$.  Below we discuss how $u_1,\ldots,u_d$ can be uncovered from $\hat{T}$.

Suppose $\tilde{T}= \sum_{i=1}^d h_i u_i$.
Set $b=au_2\ldots u_r$ if $r > 1$.  Then
\[ \widetilde{bx_1} = \sum_{i=1}^d h_i (\hat{x}_1, \hat{u}_2,\ldots,\hat{u}_r) u_i,\]
where $\hat{u}_i=(0,\ldots,1,0\ldots,0)\in k^d$ consists of all 0 except that the $i$-th coordinate is 1.
So $\widehat{bx_1}$ can be obtained from $\hat{T}$.
It is likely that $b$ generates $K$ over $k$, in which case from $\widehat{b u_j}$, $j=1,\ldots,d$, we compute the irreducible polynomial for $b$, and determine $b$ up to Galois conjugates.

 Evaluating $\widehat{bx_1}$ at $\hat{x}_1 =\widehat{bu_j}$ we obtain $\widehat{b^2 u_j}$.  Iterating we obtain $\widehat{b^i u_j}$ for $i=1,\ldots,d-1$.  From these and the irreducible polynomial of $b$ we can determine $u_j$ as a polynomial expression in $b$.  In this fashion the basis $u_1,\ldots,u_d$ can be uncovered. $\Box$

\begin{proposition}
\label{tildeF}
Given a non-constant $K$-global descent one can efficiently uncover the descent basis up to a constant factor in $K$ and a Galois conjugate.
\end{proposition}
\ \\{\bf Proof}
Let $G=F^{\sigma_i}\circ\delta^{\sigma_i}$ be a non-constant $K$-global descent for $\bu^{\sigma_i}$ with $F\in K[x_1,\ldots,x_n]$.
The proof is similar for all $i$ so assume without loss of generality $i=0$.

Write $F$ as the sum of terms $F=\sum T_i$ where $T_i = a_i m_i$ with $a_i\in K$ and $m_i$ is a monomial in $x_1,\ldots.x_n$.  Then $F\circ\delta =\sum_i a_i m_i\circ\delta$.   From $G$ we can read off $a_i m_i\circ\delta$ easily since $supp(m_i\circ\delta)\subset supp \hat{m}_i$, and $supp \hat{m}_i$ are all disjoint.  So it is enough to consider the case where $F$ is a non-constant term $T$.

Suppose for simplicity  $T=a m$ where $m=x_1\ldots x_r$ for some $r\ge 1$.  Below we discuss how $u_1,\ldots,u_d$ can be uncovered from $\hat{T}$.

We have $G= a\la \hat{m}, \bu\ra = a\delta\hat{x}_1\ldots\delta\hat{x}_r$.  Let $\hat{u}_i=(0,\ldots,1,0\ldots,0)\in k^d$ consists of all 0 except that the $i$-th coordinate is 1.  Then $\delta\hat{u}_i = u_i$. Substituting $\hat{u}_i$ for $\hat{x}_i$ in $G$ for $i=1,\ldots,r$ we obtain a polynomial
$h(\hat{x}_1) = au_2\ldots u_r \delta\hat{x}_1 =b\delta\hat{x}_1$ where $b=au_2\ldots u_r$.
Evaluating $h$ at $\hat{u}_i$ we get $bu_i$ for $i=0,\dots,d-1$.  Hence we can determine $u_i/u_0$, $i=0,\ldots,d-1$.

We remark that if $G=F\circ\delta^{\sigma_i}$ for $i > 0$, then by a similar argument we can determine $u^{\sigma_i}_j/u^{\sigma_i}_0$, $j=0,\ldots,d-1$.
$\Box$

Consider a $d$-tuple of polynomials $(f_i)_{i=0}^{d-1}$ with $f_i \in k [\hat{x}_1,\ldots,\hat{x}_n]$.   From the terms of the $d$ polynomials we can determine a set of monomials $m_1, \ldots, m_t$ in $R$ so that the support of each $f_i$ is contained in the union of the supports of $\hat{m}_1,\ldots,\hat{m}_t$.  Write $f_i = \sum_{j=1}^t f_i^{(j)}$ where the support of $f_i^{(j)}$ is contained in the support of $\hat{m}_j$, so that $(f_i)_{i=0}^{d-1} = \sum_{j=1}^t (f_i^{(j)})_{i=0}^{d-1}$.

We say that $(f_i)_{i=0}^{d-1}$ {\em contains a global descent} if there is some $j$ such that
\[ (f_i^{(j)})_{i=0}^{d-1} = \widehat{a_j m_j}\] for some $a_j\in K$.

\begin{proposition}
\label{global-descent}
 Given $(f_i)_{i=0}^{d-1}$ with $f_i\in k [\hat{x}_1,\ldots,\hat{x}_n]$ and the descent basis, one can efficiently check if $(f_i)_{i=0}^{d-1}$ contains a global descent.
\end{proposition}
\ \\{\bf Proof}
From the above discussion we are reduced to the case where there is a monomial $m$ such that $supp f_i \subset supp \hat{m}$ for all $i$.  The question is whether there is some $a\in K$ such that
$\sum_i f_i u_i = a \la \hat{m}, \bu \ra$.  This is easy to determine once we have computed $\sum_i f_i u_i$ and  $\la \hat{m}, \bu \ra$.
$\Box$

Let $G\in K[\hat{x}_1,\ldots,\hat{x}_n]$.   Let $\bt:\theta_0,\ldots,\theta^{d-1}$ be a public basis of $K/k$.
Then $G$ can be expressed in the form $G=\sum_{i=0}^{d-1} g_i \theta_i$ where $g_i\in k[x_1,\ldots,x_n]$.
By Proposition~\ref{global-descent} we can check efficiently whether $(g_i)_{i=0}^{d-1}$ contains any global descent.

Then we can write $G = \sum_i G_i$ where $supp G_i \subset supp \hat{m}_i$ for some monomial $m_i$ in $x_1,\ldots,x_n$, and $m_i$ are all distinct.

We say that $G$ {\em contains a $K$-global descent} if there is some $i$ such that $G_i$ is a $K$-global descent.
That is to say $G_i = a \la \hat{m}_i, \bu^{\sigma_j}\ra$ for some $a\in K$ and $0\le j\le d-1$.
By computing $\la \hat{m}_i, \bu^{\sigma_j}\ra$ we can determine if $G = a \la \hat{m}_i, \bu^{\sigma_j}\ra$ for some $a\in K$.

Therefore we have the following.

\begin{proposition}
\label{K-global-descent}
 Given $G\in K[\hat{x}_1,\ldots,\hat{x}_n]$, $i$ and $\bu$, one can efficiently check if $G$ contains a $K$-global descent for $\bu^{\sigma_i}$, and whether $\hat{G}_{\bt}=(g_i)_{i=0}^{d-1}$ contains any global descent, where $g_i\in k[x_1,\ldots,x_n]$ and $G=\sum_{i=0}^{d-1} g_i \theta_i = \la \hat{G}_{\bt}, \bt \ra$.
\end{proposition}

\begin{proposition}
\label{gld}
Suppose $F\in K[x_1,\ldots,x_n]$ and $\Gamma\in Gl_d (k)$.
If $\Gamma\hat{F}$ contains the global descent of a nonconstant term, then $\Gamma = \Gamma_a$ for some $a\in K$.
If $\Gamma\hat{F}=\hat{G}$ for some $G\in R$.  Then $G=aF$ for some $a\in K$ and $\Gamma = \Gamma^t_a$.  Consequently, the fraction of $\Gamma\in Gl_d(k)$ such that $\Gamma \hat{F}$ contains a global descent of a nonconstant term is negligible.
\end{proposition}
\ \\{\bf Proof}
Consider a non-constant term $T \in K[x_1,\ldots,x_n]$.  Suppose $\tilde{T}=\sum_{i=1}^d f_i u_i$.  Then for $a\in K$,
\[ \widetilde{aT} = \sum_{i=1}^d f_i au_i=\sum_j \sum_i f_i \gamma_{ij} u_j\]
where $\Gamma_a = (\gamma_{ij})$.
Hence
\[ \widehat{aT} = \Gamma_a^t \hat{T}.\]

For $a\in K$ there is a unique $b\in k^d$ such that $\delta b =\la b,\bu\ra=a$.  We denote such $b$ as $\hat{a}$. It is easy to see that $\{ \hat{T} (\hat{\alpha}) : \alpha\in K^n\}$ contains $d$ linearly independent vectors since $\widehat{T(\alpha)} = \hat{T}(\hat{\alpha})$.  Hence for $\Gamma \in Gl_d (k)$, $\Gamma \hat{T} = \hat{T}$ if and only if $\Gamma$ is the identity matrix.  It follows that $\Gamma \hat{T} = \widehat{aT}$ if and only if $\Gamma=\Gamma^t_a$.

Now let $F=\sum_i T_i$ where $T_i$ is a term.  Let $\Gamma \in Gl_d (k)$.  Then $\hat{F}=\sum_i \hat{T}_i$ and
$\Gamma\hat{F} = \sum_i \Gamma \hat{T}_i$.  If $\Gamma \hat{F}$ contains a nontrivial global descent, then
$\Gamma\hat{T}_i$ is a global descent for some $i$ where $T_i$ is a non-constant term.  This implies $\Gamma\hat{T}_i = \widehat{aT}$ for some $a\in K$.  It follows that $\Gamma =\Gamma^t_a$.

In particular if $\Gamma \hat{F}=\hat{G}$ then $G=aF$ for some $a\in K$ and $\Gamma = \Gamma_a$. $\Box$

\subsection{Specifying polynomial maps on descent varieties}
Throughout this subsection let $R=K[x_1,\ldots,x_n]$ and $\hat{R}=k[\hat{x}_1,\ldots,\hat{x}_n]$.
Suppose $V=Z(F_1,\ldots,F_m)$, the algebraic set defined by the zeroes of $F_1,\ldots,F_m \in R$.
Assume that $F_1,\ldots,\F_m$ are public.
We have $\hat{V} = Z (\hat{F}_1,\ldots,\hat{F}_m)$.  However to specify $\hat{V}$, $\hat{F}_i$ should not be used, otherwise $\bu$ may be uncovered, if $\hat{F}_i$ contains a vital term.  We choose random $\Gamma_i\in Gl_d (k)$ in secret, and check using Proposition~\ref{gld} that $\Gamma_i \hat{F}_i$ does not contain any global descent.  Let $\Gamma_i \hat{F}_i=(g_{ij})_{j=0}^{d-1}$ for $i=1,\ldots,m$.  Then $\hat{V}$ can be specified as the zero set of
$\{ g_{ij}:1\le i\le m, 0\le j \le d-1 \}$.  We have the following

\begin{proposition}
\label{specify-set}
Given $V=Z(F_1,\ldots,F_m)$ one can efficiently construct $g_{ij}\in\hat{R}$, $1\le i\le m, 0\le j \le d-1$, such that  $\hat{V}$ is the zero set of $\{ g_{ij}: 1\le i\le m, 0\le j \le d-1\}$,
$(g_{ij})_{j=0}^{d-1}$ contains no global descent and
$\Gamma_i \hat{F}_i=(g_{ij})_{j=0}^{d-1}$ for some random secret $\Gamma_i\in Gl_d (k)$, for $i=1,\ldots,m$.
\end{proposition}

Suppose a map $\varphi: V(\bar{k}) \to \bar{k}$ can be defined by the restriction of a polynomial $H\in R$ to $V$.
Then $\hat{\varphi}$ can be defined by the restriction of $\hat{H} = (h_i)_{i=0}^{d-1}$ to $\hat{V}$, with $h_i \in \hat{R}$ with coefficients in $k$.  However by Proposition~\ref{hatF}, if $H$ has a vital term then the global descent $(h_i)_{i=0}^{d-1}$ can be used to uncover the descent basis. Therefore we cannot specify $\hat{\varphi}$ by $(h_i)_{i=0}^{d-1}$.  Instead we will specify $\hat{\varphi}$ by some $(h'_i)_{i=0}^{d-1}$ where $h'_i = h_i +g_i$ with $g_i\in\hat{R}$ and $g_i$ vanishes on $\hat{V}$, so that $(h'_i )_{i=0}^{d-1}$ contains no global descent.  Simply put we want $h'_i = h_i \mod I(\hat{V})$ such that $(h'_i)_{i=0}^{d-1}$ contains  no global descent.

Similarly the map $\varphi^{\sigma_i}\circ \delta^{\sigma_i} : \hat{V}(\bar{k}) \to \bar{k}$ can be defined by the restriction of the $K$-global descent $H^{\sigma_i}\circ\delta^{\sigma_i}$ to $\hat{V}$.  We need to modify $H^{\sigma_i}\circ\delta^{\sigma_i}$ by adding a polynomial in $I(\hat{V})$ such that the resulting polynomial dose not contain any $K$-global descent.

The following propositions addresses this issue.  In the propositions we need the following assumption: Given any nontrivial monomial $m$ we can efficiently form a polynomial $F$ that vanishes on $V$ such that $m$ appears in $F$ with nonzero constant $b$.

The assumption is satisfied for example if $V = Z(F_1,\ldots,F_m)$ where $F_1$ has a nonzero constant term.   Then we can take $F=b'm_i F_1$ with random nonzero $b'\in K$.  The assumption is easy to satisfy by a linear change of coordinates.
For simplicity we also assume the degrees of $F_1$, ..., $F_m$ are bounded, as is applicable to our setting of trilinear map construction.  However we remark that the the next proposition holds when $F_i$ and $H$ are of degrees polynomially bounded.

\begin{proposition}
\label{specify-poly}
Given $H=(f_i)_{i=0}^{d-1}$ with $f_i \in\hat{R}$ of degree bounded by $O(1)$, and $A\in Gl_d (k)$, we can efficiently construct $f'_i$, $i=0,\ldots,d-1$, such that  $f'_i = f_i \mod I(\hat{V})$ and the probability that $A^t (f'_i)_{i=0}^{d-1}$ contains a global descent is negligible.
\end{proposition}

\ \\{\bf Proof}
Write $H=\sum_i H_i$ where $supp H_i \subset \hat{m}_i$, the $m_i$ are distinct monomials in $R$.
Let $\Delta$ be a vector of $d$ zero polynomials initially.
For each $H_i$, we apply the following procedure.  If $H_i =\hat{T}$ where $T=a_im_i$ with $a_i\in K$, then
choose a polynomial $F$ that vanishes on $V$ such that $m_i$ appears in $F$ with nonzero constant $b$.
Let $\Gamma$ be randomly chosen from $Gl_d (k)$.
Then $\Delta$ is replaced by $\Delta+ \Gamma\hat{F}$

After the above procedure is applied to all $H_i$, we obtain some $H+\Delta=(h'_i)_{i=1}^d$ with $h'_i\in\hat{R}$ and
$h'_i = h_i \mod I(\hat{V})$.
Moreover we have $(h'_i)_{i=1}^d = \sum_i H'_i$ where $H'_i \in \hat{R}^d$, and each $H'_i$ is of the form
$H'_i=G_i+\sum_j \Gamma_{ij} \hat{m}_i$ where $m_i$ is a monomial, $G_i$ ie either 0 or $G_i=H_i$, $supp H_i\subset supp\hat{m}_i$, and $\Gamma_{ij}$ are random elements in $Gl_d (k)$.
Let $A\in Gl_d (k)$.  If $A^t H'_i = \widehat{am_i}$ for some $a\in K$, then $A^t G_i +A^t(\sum_j \Gamma_{ij})\hat{m}_i = \Gamma^t_a\hat{m}_i$ for some $a\in K$.  Hence $G_i = B \hat{m}_i$ where $B$ is in the additive group generated by $Gl_d (k)$.  However when $G_i = B \hat{m}_i$ where $B$ is in the additive group generated by $Gl_d (k)$, it is unlikely  $A^t(B+\sum_j \Gamma_{ij}) = \Gamma^t_a$ for some $a\in K$ with random $\Gamma_{ij}$.
Consequently it is unlikely that $A^t H'_i = \widehat{am_i}$ for some $a\in K$.
$\Box$

\subsection{Specifying rational maps on descent varieties}
We keep the same notation as before, but suppose now the map $\varphi: V(\bar{k}) \to \bar{k}$ can be defined by the restriction of a rational function $F/G$ to $V$ where $F,G\in R$.  Then $\hat{\varphi}:\hat{V}\to\bar{k}^d$ is the descent map defined by $\hat{\varphi} = \rho^{-1}\circ \prod_{i=0}^{d-1} \varphi^{\sigma_i} \circ \rho$.  We have
$\delta^{\sigma_i}\circ\hat{\varphi} = \varphi^{\sigma_i}\circ\delta^{\sigma_i}$, or more explicitly,
\[ \la \hat{\varphi} (\hat{x}), \bu^{\sigma_i}\ra= \varphi^{\sigma_i}(\delta^{\sigma_i}\hat{x})\]
for all $\hat{x}\in\hat{V}(\bar{k})$.

\begin{lemma}
\label{aF}
Let $F\in K[x_1,\ldots,x_n]$, and $a\in K^*$ and $i\in\{0,\dots,d-1\}$.
Then
\begin{enumerate}
\item $\la \widehat{aF}, \bu^{\sigma_i} \ra = a^{\sigma_i} \la \hat{F}, \bu^{\sigma_i} \ra$.
\item $\widehat{aF} = A^t \hat{F}$ where $A\in Gl_d (k)$ such that $a\bu = A \bu$.
\end{enumerate}
\end{lemma}
\ \\{\bf Proof}
\[ \la \widehat{aF}, \bu^{\sigma_i} \ra = (aF)^{\sigma_i} \circ \delta^{\sigma_i} = a^{\sigma_i} (F^{\sigma_i}\circ\delta^{\sigma_i}) = a^{\sigma_i} \la \hat{F}, \bu^{\sigma_i} \ra.\]  Hence the first assertion.
Since $A$ is fixed by $\sigma$, $a\bu = A\bu$ implies $a^{\sigma_i} \bu^{\sigma_i} = A \bu^{\sigma_i}$.
\[a^{\sigma_i} \la \hat{F}, \bu^{\sigma_i} \ra=\la \hat{F}, a^{\sigma_i}\bu^{\sigma_i}\ra =\la \hat{F}, A\bu^{\sigma_i}\ra=\la A^t \hat{F},\bu^{\sigma_i}\ra.\]
Hence the second assertion, given that $\rho$ is invertible.. $\Box$

For $i=0,\ldots,d-1$ let $\Gamma_{\bt,i}$ denote the matrix in $Gl_d (k)$ such that $\bu^{\sigma_i} = \Gamma_{\bt,i} \bt$.

For $a\in K$, let $\Gamma_{\bt,a}$ denote the matrix in $Gl_d (k)$ such that $a \bu = \Gamma_{\bt,a} \bt$.

\begin{lemma}
\label{ai}
For $a\in K$ and $i=0,\ldots,d-1$
\[ a\la \hat{x}, \bu^{\sigma_i}\ra = \la \Gamma^t_{\bt,a}\Gamma^t_{\bt,i} \hat{x},\bt\ra\]
for all $\hat{x}\in \bar{k}^d$.
\end{lemma}
\ \\{\bf Proof} \[ a\la \hat{x}, \bu^{\sigma_i}\ra = a\la \hat{x}, \Gamma_{\bt,i}\bt\ra=
a\la \Gamma^t_{\bt,i} \hat{x}, \bt\ra=
\la \Gamma^t_{\bt,i} \hat{x}, a\bt\ra=\la \Gamma^t_{\bt,i} \hat{x}, \Gamma_{\bt,a}\bt\ra=
\la \Gamma^t_{\bt,a}\Gamma^t_{\bt,i} \hat{x},\bt\ra\] $\Box$

\begin{proposition}
\label{avarphi}
Suppose $\varphi (x) = F(x)/G(x)$ for all $x\in V(\bar{k})$.  Then for $a\in K$ and $i=\ldots,d-1$, the following holds.
\begin{enumerate}
\item
For all $r\in K^*$,
\[ a \varphi^{\sigma_i} (\delta^{\sigma_i} \hat{x}) = \frac{F_1^{\sigma_i}\circ\delta^{\sigma_i}} {G_1^{\sigma_i}\circ\delta^{\sigma_i}}=\frac{\la \hat{F}_1, \bu^{\sigma_i} \ra}{\la \hat{G}_1,\bu^{\sigma_i}\ra}\]
where $F_1 = (ar)^{\sigma_{-i}} F$ and $G_1 = (r)^{\sigma_{-i}} G$.
\item
The function $a\varphi\circ\delta^{\sigma_i}$ on $\hat{V}$ can be defined by $\frac{\sum_{i=0}^{d-1} f'_i\theta_i}{\sum_{i=0}^{d-1} g'_i\theta_i}$ where $f'_i,g'_i\in k[\hat{x}_1,\ldots,\hat{x}_n]$,
both $(f'_i)$ and $(g'_i)$ contain no global descent and both functions $\sum_{i=0}^{d-1} f'_i\theta_i$ and $\sum_{i=0}^{d-1} g'_i\theta_i$ contain no $K$-global descent.
\end{enumerate}

\end{proposition}
\ \\{\bf Proof} We have
\[ a\varphi^{\sigma_i} (\delta^{\sigma_i} \hat{x}) = a \frac{F^{\sigma_i} (\delta^{\sigma_i} \hat{x})}{G^{\sigma_i} (\delta^{\sigma_i} \hat{x})} = a \frac{\la \hat{F}(\hat{x}), \bu^{\sigma_i}\ra}{\la \hat{G}(\hat{x}), \bu^{\sigma_i}\ra} = \frac{ar \la  \hat{F}(\hat{x}), \bu^{\sigma_i}\ra}{r\la \hat{G}(\hat{x}), \bu^{\sigma_i}\ra}. \]
Then the first assertion follows from Lemma~\ref{aF}.

Applying Proposition~\ref{specify-poly} to $\hat{F}_1$ we construct efficiently $F'_1$, a $d$-tuple of polynomials in $k[\hat{x}_1,\ldots,\hat{x}_n]$, such that $F'_1 -\hat{F}_1$ is a $d$-tuple of polynomials in $I(\hat{V})$, and
$\Gamma_{\bt,i}^t F'_1$, and $A_j^t F'_1$ all contain no global descent for $j=0,\ldots,d-1$, where $A_j=\Gamma_{\bt,i}\Gamma^{-1}_{\bt,j}$.
Let $(f'_j)_{j=0}^{d-1} = \Gamma_{\bt,i}^t F'_1$.  Then $\sum_{j=0}^{d-1} f'_j \theta_j = \la\hat{F}, \bu^{\sigma_i}\ra$ on $\hat{V}$.
Moreover,
\[ \sum_{j=0}^{d-1} f'_j \theta_j = \la (f_j)_{j=0}^{d-1}, \bt\ra = \la \Gamma_{\bt,j}^{-t} (f_r)_{r=0}^{d-1}, \bu^{\sigma_j}\ra=\la A_j^t F'_1, \bu^{\sigma_j}\ra.\]
Consequently $\sum_{j=0}^{d-1} f'_j \theta_j$  contains no $K$-global descent for $\bu^{\sigma_j}$ for all $j$.

The tuple $(g'_i)$ can be constructed similarly by applying Proposition~\ref{specify-poly} to $\hat{G}_1$.
$\Box$

By choosing $r\in K^*$ in Proposition~\ref{avarphi} randomly, the coefficient $a$ is blinded.
When $a \varphi^{\sigma_i} \circ \delta^{\sigma_i}$ is specified in the form $\frac{\sum_i f'_i \theta_i}{\sum_i g'_i\theta_i}$ as in Proposition~\ref{avarphi}, we say that the specification contains no global descent and the coefficient $a$ is blinded.

To specify $\hat{\varphi}$, we observe that $\hat{\varphi}=(\hat{\varphi}_i)_{i=0}^{d-1}$ where
\[\hat{\varphi}_i (\hat{x}) = \sum_{j=0}^{d-1} w_{ij} \varphi^{\sigma_i}(\delta^{\sigma_i} \hat{x}).\]
Let $h_{ij}=w_{ij} \varphi^{\sigma_j} \circ \delta^{\sigma_j}$.  For each $i$,
partition the set $\{h_{ij}:j=0,\ldots,d-1\}$ into random disjoint subsets $S_0$, $S_1$,..., where each $S_j$ consists of 2 or 3 functions.  Then $\hat{\varphi}_i = \sum_j \psi_{ij}$, where $\psi_{ij}$ is the sum of the functions in $S_j$.  The function $h_{ij}=w_{ij} \varphi^{\sigma_j} \circ \delta^{\sigma_j}$ can be expressed in the quotient form as in Proposition~\ref{avarphi}, so that the expression contains no global descent.  As we take the sum of the functions in $S_j$ and express the resulting function $\psi_{ij}$ in quotient form again, we can apply Proposition~\ref{specify-poly} to modify each tuple of polynomials if necessary and make sure that the expression, which is to be used to specify $\psi_{ij}$,  does not contain any global descent.  Therefore we have the following

\begin{proposition}
\label{specify-rat}
To specify $\hat{\varphi}=(\hat{\varphi}_i)_{i=0}^{d-1}$, where $\hat{\varphi}_i=\sum_j h_{ij}$, with $h_{ij}=w_{ij} \varphi^{\sigma_j} \circ \delta^{\sigma_j}$, one can specify $\hat{\varphi}_i$ as
 $\sum_j \psi_{ij}$ where $\psi_{ij}$ is the sum of the functions in $S_{ij}$. Each $S_{ij}$ contains 2 or 3 functions and for each $i$, $S_{ij}$ form a random partition of $\{ h_{ij}: j=0,\ldots,d-1\}$ into subsets of 2 or 3 elements.  Moreover $\psi_{ij}$ can be efficiently specified in the form $\frac{\sum_{i=0}^{d-1} f'_i\theta_i}{\sum_{i=0}^{d-1} g'_i\theta_i}$ where $f'_i,g'_i\in k[\hat{x}_1,\ldots,\hat{x}_n]$,
both $(f'_i)$ and $(g'_i)$ contain no global descent and both functions $\sum_{i=0}^{d-1} f'_i\theta_i$ and $\sum_{i=0}^{d-1} g'_i\theta_i$ contain no $K$-global descent.
\end{proposition}

\subsection{Linear analysis}
\label{linear-analysis}

Suppose the descent $\hat{\varphi}$ of a map $\varphi: V\to V$ defined over $K$ is specified.
Suppose one point on $\hat{V}$ is given. Then  starting with the given point, one can repeatedly apply the descent map $\hat{\varphi}$ to obtain more points on $\hat{V}$.   Heuristically speaking we may consider these points as randomly sampled from $\hat{V} (\bar{k})$.

In this section we investigate the following question:  under what conditions could information about $\bu$ be efficiently computed from  polynomially many sampled points $\alpha$ of $\hat{V}$?

Similarly, suppose $\phi:V\to \bar{k}$ is a rational function defined over $K$, and suppose $\phi$ is not public but
$\phi\circ\delta :\hat{V}\to \bar{k}$ is specified, so that polynomially many pairs of $(\alpha,\phi\circ\delta (\alpha))$ can be obtained where $\alpha\in\hat{V}$.
We also investigate the following question: under what conditions could information about $\bu$ be efficiently computed from polynomially many pairs $(\alpha,\phi\circ\delta (\alpha))$ (even if
$\hat{\varphi}$ and $\phi\circ\delta$ are specified properly so that the specifications contain no global descent and none of the sampled points are descent points)?

We begin with some general consideration and definitions.  Suppose $S$ is a finite set of monomials in variables $x_1,\ldots,x_n$.  Let $\lambda_S$ denote the map from $\bar{k}^n\to \bar{k}^{|S|}$ such that for $\alpha\in \bar{k}^n$, $\lambda_S (\alpha)$ is the vector consisting of $m(\alpha)$ where $m$ ranges over all monomials in $S$.

For $A\subset \bar{k}^n$, let $\ell_S (A)$ be the dimension of the linear space of $\{ F\in \bar{k}[x_1,\ldots,x_n]: supp F\subset S, F(\alpha)=0, \forall\alpha\in A\}$, and let $\omega_S (A)$ be the maximal number of linearly independent $\lambda_S (\alpha)$ with $\alpha\in A$.

Let $F\in\bar{k}[x_1,\ldots,x_n]$ with $supp F\subset S$.  Write $F = \sum_{m\in S} c_m m$.  Then for $\alpha\in\bar{k}^n$, $F(\alpha) = \la c_F, \lambda_S (\alpha)\ra$ where $c_F$ is the vector consisting of $c_m$ with $m\in S$.  Therefore, $\ell_S (A)+ \omega_S (A)=|S|$.

If $\omega_S (A)=|S|$, then $F$ is the unique polynomial $G$ with support contained in $supp G\subset S$ such that $G(\alpha)= F(\alpha)$ for all $\alpha\in A$.
If $\omega_S (A)=|S|-1$, then there is a unique non-zero polynomial $F$ up to a constant multiple such that $F(\alpha)=0$ for all $\alpha\in A$.

We now consider as before, the situation of an algebraic variety $V$ defined by a set of polynomials in $K[x_1,\ldots,x_n]$ and its descent $\hat{V}$ with respect to a basis $\bu$ of $K$ over $k$.  Suppose a set $A$ of polynomially many sampled points on $\hat{V}$ is available.

Suppose $S$ is a set of monomials in $x_1,\ldots,x_n$.  Let $\hat{S} = \cup_{m\in S} supp\hat{m}$.
Let $I(V)$ be the ideal consisting of polynomials in $K[x_1,\ldots,x_n]$ that vanish at all points of $V$.
Let $I_S$ be the set of polynomials in $I(V)$ with support bounded by $S$.
Let $L_{\hat{S}}$ be the linear space of polynomials in $K[\hat{x}_1,\ldots,\hat{x}_n]$ with support bounded by $\hat{S}$ that vanish at all the sampled points of $\hat{V}$.  Hence $\ell_{\hat{S}} (A) =\dim L_{\hat{S}}$.

\begin{lemma}
\label{linear-attack}
Suppose $\varphi: \hat{V}\to \bar{k}$ is a map such that there is some $F\in K[x_1,\dots,x_n]$ and $\varphi$ can be defined by the restriction of $F\circ\delta$ to $\hat{V}$.  Let $S=supp F$.  If $\ell_{\hat{S}} (A) =0$ then
$F\circ\delta$ is uniquely determined from $\lambda_{\hat{S}} (A)$ and $\varphi (A)$.
\end{lemma}

Since  $F\circ\delta$ is a $K$-global descent, it reveals substantial information on $\bu$ by Proposition~\ref{tildeF}.
Lemma~\ref{linear-attack} leads to the following attack.   We need to assume $n=O(1)$.  Suppose a map $\varphi: \hat{V}\to\bar{k}$ is specified in some way but it can actually be defined as $F\circ\delta$ for some $F\in K[x_1,\ldots,x_n]$ of bounded degree.   Suppose $S=supp F$. Since we assume $\deg F$ is bounded, there are only finitely many choices for $S$, hence $\hat{S}$.

Suppose the correct $S$ is being tried.  If $\ell_{\hat{S}} (A)=0$,  then $F\circ\delta$ is the unique polynomial with support bounded by $\hat{S}$ such that $F\circ\delta (\alpha) = \varphi (\alpha)$ for all $\alpha\in A$.   From $\varphi(A)=\{\varphi(\alpha): \alpha\in A\}$ and $\lambda_{\hat{S}} (A)$ the coefficient vector of $F\circ\delta$ can be determined, hence $F\circ\delta$ is found.

Consider for example when $F=x_1$.  Suppose the projection of $V$ to the $x_1$-coordinate is surjective.  Then the projection of $\hat{V}$ to the coordinates in $\hat{x}_1$ is also surjective.  We have $supp F = \{ x_1\}$.
In this situation it is likely that $\omega_{\hat{S}} (A) = d = |\hat{S}|$, and the attack described above can be mounted.

The attack can be avoided if we make sure that whenever some $\varphi:\hat{V}\to \bar{k}$ is specified and $\varphi$ can be defined as the restriction of $F\circ\delta$ on $\hat{V}$, $I_S\neq 0$ where $S=supp F$.  More precisely suppose $h\in I_S$.
Then $supp \hat{h}\subset \hat{S}$, so $\ell_{\hat{S}} (A) > 0$.
Letting $\hat{h}=(h_i)_{i=0}^{d-1}$, we know that $F\circ\delta + \sum a_i h_i$ defines the same function on $\hat{V}$ for all $a_i\in K$.  Moreover $supp\hat{h}\subset supp\hat{F}=supp F\circ\delta$.  It follows from Lemma~\ref{kgld} and Proposition~\ref{gld} that for random choices of $a_i$, the probability that $\sum_i a_i h_i$ is a $K$-global descent is negligible, hence the probability that  $F\circ\delta + \sum a_i h_i$ is a $K$-global descent is negligible.   We have the following:

\begin{lemma}
\label{no-linear-attack}
Let $F\in K[x_1,\ldots,x_n]$ with $S=supp F$.
Suppose $I_S\neq 0$.  Let $h\in I_S$ with $\hat{h}=(h_i)_{i=0}^{d-1}$.  Then $F\circ\delta + \sum a_i h_i$ defines the same function on $\hat{V}$ for all $a_i\in K$. Moreover for random choices of $a_i$, the probability that  $F\circ\delta + \sum a_i h_i$ is a $K$-global descent is negligible.
\end{lemma}

The attack described below, {\em linear-term attack}, though very limited in scope of success, helps identify some relatively weak cases, such as when $V$ is contained in a hyperplane, or
when $V$ is defined by a single polynomial with a linear term.

\begin{lemma}
\label{linear}
Suppose there is some $F\in I(V)$ with a linear term $x_i$ and $\ell_{\hat{S}} (A) = d$ where
$S=supp F$.  Then from $S$ and $A$, the set of sampled points, $\hat{F}$ can be uncovered efficiently.
\end{lemma}
\ \\{\bf Proof}  Since $S=supp F$, the polynomials in $\hat{F}$ all have support contained in $\hat{S}$.  Suppose without loss of generality $F = x_1 + F_1$ where $x_1\not\in supp F_1$.
Then $\hat{F}= \hat{x}_1 + \hat{F}_1$.  If we put $\hat{F}=(f_i)_{i=0}^{d-1}$ then
$f_i = x_{1i} + g_{i}$ where $\hat{F}_1 = (g_i)_{i=0}^{d-1}$.  So $x_{1j}$ is not in $supp g_i$ for $j\neq i$.
It follows that $f_0,\ldots,f_{d-1}$ are linearly independent.  If $\ell_{\hat{S}} (A)= d$, then $f_0,\ldots,f_{d-1}$ form a linear basis for the linear space $L$ of polynomials $G$ with support bounded by $\hat{S}$ such that $\la C_G, \lambda_{\hat{S}} (\alpha) \ra=0$ for all $\alpha\in A$ where $C_G$ denotes the coefficient vector of $G$.  Moreover $f_i$ can be found by solving for $f\in L$ such that the coefficient of $f$ in $x_{1i}=1$ and the coefficient of $f$ in $x_{1j}=0$ for $j\neq i$.  $\Box$

\ \\{\bf Example 1} Suppose $V$ is contained in a hyperplane defined by a linear polynomial $H$.
Let $S=supp H$.  Then $\dim L_{\hat{S}} \ge d$, and $L_{\hat{S}}$ contains all the linear polynomials in $\hat{H}$.
If $\dim L_{\hat{S}} = d$ then $\hat{H}$ can be determined from $S$ and the sampled points by Lemma~\ref{linear}.

\ \\{\bf Example 2} Let $V$ be the affine part of the elliptic curve defined by $y^2=x^3+ax+b$.  Consider $S=\{ y^2, x^3, x, 1\}$, and $F=a^{-1} (y^2 -x^3 -ax-b)$.  If $\dim L_{\hat{S}} =d$ then $\hat{F}$ can be can be determined from $S$ and the sampled points by Lemma~\ref{linear}.  The situation is similar if $V$ is defined by a polynomial that contains a linear term.

Suppose $F\in I(V)$ with a linear term $x_i$.
To prevent linear-term attack to discover $\hat{F}$, it is sufficient if  $I_{S'} \neq 0$ where $S'=S-\{ x_i\}$
and $S=supp F$.  Suppose  $h\in I_{S'}$.  Let $\hat{h}=(h_i)_{i=0}^{d-1}$.  Then $h_i \in L_{\hat{S'}}\subset L_{\hat{S}}$ for all $i$.  It follows that $\dim  L_{\hat{S}} > d$.  To summarize we have the following:
\begin{lemma}
\label{no-linear}
Suppose $F\in I(V)$ with a linear term $x_i$.  Let
$S=supp F$.  If $I_{S'} \neq 0$ where $S'=S-\{ x_i\}$, then $\ell_{\hat{S}} (A) > d$.
\end{lemma}

\subsection{Choosing a birational model to prevent linear attacks}
\label{birational-model}
To prevent the linear attacks described in the previous subsection, we can form $V'$ birational to $V$ over $K$ such that
conditions preventing the attacks as described in Lemma~\ref{no-linear-attack} and Lemma~\ref{no-linear} can be easily satisfied.

Consider a rational map $\lambda:\bar{k}^n \to \bar{k}^{n+1}$ sending $(x_1,\ldots,x_n)$ to
$(x_1,\ldots,x_n,x_{n+1})$ where
\[ \sum_{1\le i < j \le n} a_{ij} x_i x_j + x_{n+1} \sum_{i=1}^n b_i x_i=0\]
where $a_{ij}\in K$ for all $i,j$ and $b_i\in K$ for all $i$.   The map $\lambda$ is injective where $\sum_{i=1}^n b_i x_i \neq 0$.  We assume $b_i$ are randomly chosen from $K$, so that with high probability no sampled points lie on the exceptional hyperplane $\sum_{i=1}^n b_i x_i =0$ .

Consider a random general linear map $\mu: \bar{k}^{n+1}\to \bar{k}^{n+1}$ given by an $n+1$ by $n+1$ invertible matrix over $B$ over $K$.  Let $L_1$,...$L_{n+1}$ be linear forms in $x_1,\ldots,x_{n+1}$ such that letting $x =(x_1,\ldots,x_{n+1})$, then $\mu^{-1} (x) = (L_i (x))_{i=1}^{n+1}$.

Let $V$ be an algebraic variety defined by a set of polynomials $F_1,\ldots,F_m$ in $K[x_1,\ldots,x_n]$ as before.  Let $V'=\mu(\lambda(V))$, the image of $V$ under $\iota=\mu\circ\lambda$.  Then $\iota:V\to V'$ is $K$-birational and $V'$ is the zero set of
$F'_i (x)= F_i (L_1 (x),\ldots, L_n (x))$, $i=1,\ldots,m$, and
$R'= \sum_{1\le i < j \le n} a_{ij} L_i (x) L_j (x) + L_{n+1} (x) \sum_{i=1}^n b_i L_i (x)$.

Note that $L_i L_j$ likely involves all $x_r x_s$ with randomly chosen $\mu$, and more generally,
$L_{i_1}\ldots L_{i_s}$ likely involves all monomials in $x_1,\ldots,x_{n+1}$ of degree $s$.  Suppose $F\in K[x_1,\ldots,x_n]$.  Then $F(L_1 (x),\ldots, L_n (x))$ is likely dense in $x_1,\ldots,x_{n+1}$.  This is useful in preventing the attacks described in Lemma~\ref{linear-attack} and Lemma~\ref{linear} as we explain below.

We assume that $I(V)$ does not contain any linear polynomial.  In forming $V'$ we assume that with randomly chosen $\mu$ that every polynomial in the defining set  of $V'$ is dense at least for some degree $i > 1$, in the sense that the support of the polynomial contains all monomials in $x_1,\ldots,x_{n+1}$ of degree $i$.  Let $I' = I(V')$ and $I'_S$ denote the subset of $I'$ with support bounded by $S$ where $S$ is a set of monomials in $x_1,\ldots,x_{n+1}$.  If $S$ contains all monomials of degree $i$ for some $i\ge 2$, then
$I'_S\neq 0$ as it contains for example $R' x_1^{i-2}$. Hence the attack in Lemma~\ref{linear} can be prevented if every polynomial in $I'$ of bounded degree with a linear term is dense for some degree at least 2.

Suppose $\varphi:V\to \bar{k}$ is a $K$-rational function that can be defined by $F/G$ on $V$ where $F,G\in K[x_1,\ldots,x_n]$ both of degree at least 2.  Then on $V'$ the corresponding map $\varphi'=\varphi\circ\iota^{-1}$ can be defined by
$F'/G'$ with $F',G'\in K[x_1,\ldots,x_{n+1}]$ where $F'=F(L_1 (x),\ldots, L_n (x))$ and $G'=G(L_1 (x),\ldots, L_n (x))$, and both $F'$ and $G'$ are likely dense.
Suppose $F'$ (resp. $G'$) is dense for degree $i \ge 2$.  Then $I'_S \neq 0$ where $S=supp F'$ (resp. $S=supp G'$).  Hence the attack in Lemma~\ref{linear-attack} can be prevented when $\varphi'\circ\delta$ is specified by specifying
$F'\circ\delta$ and $G'\circ\delta$.

Suppose $\varphi:V\to V$ is a $K$-rational map.  The map $\iota\circ\varphi:V\to V'$ consists of $n+1$ rational functions
$\psi_1$,...,$\psi_{n+1}$.  Suppose $\psi_i$ can be defined by $F_i/G_i$ with $F_i,G_i\in K[x_1,\ldots,x_n]$.  Then
$\varphi' = \iota\circ\varphi\circ\iota^{-1}: V'\to V'$ can be defined by $\psi'_i=F'_i/G'_i$ where
$F'_i = F_i (L_1(x),\ldots,L_n (x))$ and  $G'_i = G_i (L_1(x),\ldots,L_n (x))$  for $i=1,\ldots,n+1$.  Similarly, if $F'_i$ and $G'_i$  are dense in some degree at least 2, then the attack in Lemma~\ref{linear-attack} can be prevented when $F'_i\circ\delta$ and $G'_i\circ\delta$ (and their Galois conjugates) are specified in order to specify $\hat{\varphi'}$.

In our situation there will be a finite set of polynomials involved in defining various rational functions on $V$ that are of interest to trilinear map construction.   By choosing random $\mu$ we check and make sure that every such polynomial $F$ is such that the corresponding $F'=F(L_1 (x),\ldots, L_n (x))$ is dense for some degree $i \ge 2$.

\subsection{Summary on specification of descent maps and functions}
\label{sum}
We now summarize our discussion from \S~\ref{gd} to \S~\ref{birational-model}.

We say that an algebraic set $V$ defined over $K$ is {\em safe} (for specification of the descent of $V$) if the following holds: (1) $I(V)$ contains no linear polynomial, and (2) if $F\in I(V)$ contains a linear term $x_i$ then $I_S \neq 0$ where $S=supp F - \{x_i \}$.
Condition (1) is easy to satisfy unless $V$ is a linear variety.  Condition (2) is likely to hold after a random birational transformation as described in \S~\ref{birational-model}.  These conditions prevent the linear-term attack described in Lemma~\ref{linear} using the sampled points on $\hat{V}$.

For $F\in K[x_1,\ldots,x_n]$, let $S_F$ be the support of $F$.  We say that $F$ is {\em safe} if $I_{S_F}\neq 0$.
When $F$ is safe the linear attack (Lemma~\ref{linear-attack}) does not apply when $F^{\sigma_i}\circ\delta^{\sigma_i}$ is specified (Lemma~\ref{no-linear-attack}).

A rational function $\varphi: V\to \bar{k}$ defined over $K$ is {\em safe} if for $F,G\in K[x_1,\ldots,x_n]$ such that $\varphi$ can be defined by $F/G$ on $V$,  $F$ and $G$ are safe.

When a random birational transformation $\iota$ as described in \S~\ref{birational-model} is applied to $V$, the rational function $\varphi\circ\iota^{-1}$ which replaces $\varphi$ is likely safe if $\varphi$ is defined as the quotient of two polynomials of degree at least 2.

The specification of descent maps and descent functions with blinding multiples ($\hat{\varphi}$ and $a\varphi^{\sigma_i}\circ\delta^{\sigma_i}$) all boil down to specifying functions on $\hat{V}\to \bar{k}$ of the form
$r F^{\sigma_i}\circ\delta^{\sigma_i}$ where $r$ is secret random and $F\in K[x_1,\ldots,x_n]$ may be known.  By Proposition~\ref{avarphi}, $r F^{\sigma_i}\circ\delta^{\sigma_i}$ can be specified as $\sum_{i=0}^{d-1} f_i \theta_i$ with $f_i \in k[\hat{x}_1,\ldots,\hat{x}_n]$ such that $(f_i)_{i=0}^{d-1}$ contains no global descent and $\sum_{i=0}^{d-1} f_i \theta_i$ contains no $K$-global descent.  We say that $r F^{\sigma_i}\circ\delta^{\sigma_i}$ is {\em properly specified} (as $\sum_{i=0}^{d-1} f_i \theta_i$).

The map $a\varphi^{\sigma_i} \circ \delta^{\sigma_i}$ is specified once we specify $ar F^{\sigma_i}\circ\delta^{\sigma_i}$ and $rG^{\sigma_i}\circ\delta^{\sigma_i}$ where $r\in K^*$ is secret and randomly chosen.
We say that $a\varphi^{\sigma_i} \circ \delta^{\sigma_i}$ is {\em properly specified} if  $ar F^{\sigma_i}\circ\delta^{\sigma_i}$ and $rG^{\sigma_i}\circ\delta^{\sigma_i}$ are properly specified, where $r\in K^*$ is secret and randomly chosen, such that $\varphi = F/G$ on $V$, we say that $a$ is {\em blinded} in the specification.

When the descent map $\hat{\varphi}$ is specified in the manner as described in Proposition~\ref{specify-rat}, we say that it is {\em properly specified}.

\subsection{Mixed descent}
\label{mixed-descent}
Let $\varphi : V\times V \to \bar{k}$ be a rational function that can be defined by $F(x,y)/G(x,y)$ where
$F$ and $G$ are polynomials in $x=x_1,\ldots,x_n$ and $y=y_1,\ldots,y_n$.  We now consider descent function determined by $\varphi$ on $\hat{V}\times\hat{V'}$ where $\hat{V}$ and $\hat{V'}$ are descent varieties of $V$ formed with respect two different secret bases. We discuss how the method for properly specifying descent functions can be naturally adapted to this situation.

In this setting, we
fix a public basis $\theta_1,\ldots,\theta_d$ of $K/k$, a private basis $\bu=u_0,\ldots,u_{d-1}$ of $K/k$, and another private basis $\bu'=u'_0,\ldots,u'_{d-1}$ of $K/k$.

Let $\delta$ denote the basic descent map $\bar{k}^d \to \bar{k}$ with respect to $\bu$, and $\rho$ the bijective linear map $\bar{k}^d \to \bar{k}^d$ determined by $\delta$.

Let $\delta'$ denote the basic descent map $\bar{k}^d \to \bar{k}$ with respect to $\bu'$, and $\rho'$ the bijective linear map $\bar{k}^d \to \bar{k}^d$ determined by $\delta'$.

Let $\hat{V}$ denote the descent of $V$ with respect to the basis $\bu$.

Let $\hat{V}'$ denote the descent of $V$ with respect to the basis $\bu'$.

Suppose $A$ is the set of sampled points on $\hat{V}$ and $A'$ is the set of sampled points on $\hat{V'}$.

Then $\varphi\circ (\delta^{\sigma_i}, {\delta'}^{\sigma_i})$ is a descent function $\hat{V}\times\hat{V'}\to \bar{k}$ that can be defined by $F'/G'$ where $F'=F^{\sigma_i} (\delta^{\sigma_i} \hat{x}, {\delta'}^{\sigma_i} \hat{y} )$
and $G'=G^{\sigma_i} (\delta^{\sigma_i} \hat{x}, {\delta'}^{\sigma_i} \hat{y} )$.

We say that a polynomial $F(x,y)$ is {\em safe} if  $I_{S_1 (F)} \neq 0$  where
$S_1 (F)$ is the support of $F(x,y)$ as a polynomial in $x$, and $I_{S_2 (F)} \neq 0$  where
$S_2 (F)$ is the support of $F(x,y)$ as a polynomial in $y$.

Suppose $F$ is safe.  We expect $F(x,\delta'\beta)$ to be safe for randomly sampled $\beta$ from $\hat{V'}$.  Similarly we expect $F(\delta\alpha, y)$ to be safe for randomly sampled $\alpha$ from $\hat{V}$.

We say a rational function $\varphi : V\times V \to \bar{k}$ defined over $K$ is {\em safe}, if for $F,G\in K[x,y]$ such that $\varphi$ can be defined by $F/G$ on $V\times V$, both $F$ and $G$ are safe.

Performing a birational transformation as in \S~\ref{birational-model} if necessary we may assume $F$ and $G$ are dense in both $x$ and $y$, hence the above conditions are likely to hold for $F$ and $G$ of degree at least 2 in $x$ and in $y$.

We discuss how $F'$ can be properly specified as a function $\hat{V}\times\hat{V'}\to \bar{k}$.  The method can also be applied to $G'$.
We consider the case $F'=F(\delta\hat{x},\delta\hat{y})$ in the discussion below.  The general case $F'=F^{\sigma_i} (\delta^{\sigma_i} \hat{x}, {\delta'}^{\sigma_i} \hat{y} )$ can be treated in a similar fashion.
So suppose $F(x,y)=\sum_i a_i m_i m'_i$ where $a_i\in K^*$, $m_i$ is a monomial in $x=x_1,\ldots,x_n$ and $m'_i$ is a monomial in $y=y_1,\ldots,y_n$.  Then
$F'(\hat{x},\hat{y}) = F(\delta\hat{x}, \delta\hat{y})=\sum_i a_i m_i(\delta\hat{x}) m'_i(\delta'\hat{y})$, a mixed $K$-global descent with respect to $\bu$ and $\bu'$.  From this form of $F'$ one easily obtains $a_i m_i(\delta\hat{x}) m'_i(\delta'\hat{y})$, from which essential information on $\bu$ may be obtained by specializing $\hat{y}$ to random
$\beta\in A'$, similarly essential information on $\bu'$ may be obtained by specializing $\hat{x}$ to random
$\alpha\in A$.  So in specifying the function on $\hat{V}\times\hat{V'}$ we want to modify $F'$ into some $H$ where $H-F'$ vanishes on $\hat{V}\times\hat{V'}$ and $H$ does not contain any $K$-global descent with respect to $\bu$ or $\bu'$ even after specialization at sampled points.  This can be achieved by adapting the method described in the proof of Proposition~\ref{specify-poly}.

Let $F'=\sum_i a_i m_i(\delta\hat{x}) m'_i(\delta'\hat{y})=\sum_i M_i M'_i$ where
$M_i = \la A_i^t \hat{m}_i, \bu\ra$ with $A_i = \Gamma_{ar^{-1}}$ with $r\in K^*$ being randomly chosen and $\Gamma_{ar^{-1}} \bu = ar^{-1} \bu$, and
$M'_i = \la B_i^t \hat{m'}_i, \bu'\ra$ with $B_i = {\Gamma'}_{r}$ and ${\Gamma'}_r \bu' = r\bu'$.

As in Proposition~\ref{specify-poly} we
find polynomials $F_i(x)\in I(V)$ containing a term $m_i$ and $F'_i(y)\in I(V)$ containing a term $m'_i$.  We
modify the polynomial $\sum_i M_i M'_i$ to $\sum_i H_i H'_i$ where
\[ H_i = M_i + \la \Gamma^t\hat{F}_i,\bu\ra,\]
\[ H'_i = M'_i + \la \Gamma'^t\hat{F'}_i,\bu\ra\]
with randomly chosen $\Gamma,\Gamma'\in Gl_d (k)$.

After the modification the polynomial takes the form $\sum_i N_i N'_i$ with
$N_i = \la \Gamma^t_i \hat{m}_i, \bu\ra$ where $m_i$ is a monomial in $x$, $\Gamma_i$ is a random sum of matrices
in $Gl_d (k)$ except for at most one matrix of the form $\Gamma_a$ where $a\in K^*$ and $\Gamma_a \bu = a\bu$.  Hence
$\Gamma_i$ is most likely not of the form $\Gamma_b$ for some $b\in K^*$, in which case $N_i$ is not a $K$-global descent with respect to $\bu$.  Similarly $N_i$ is unlikely a $K$-global descent with respect to $\bu^{\sigma_j}$ and ${\bu'}^{\sigma_j}$ for $j=0,\ldots,d-1$.  For example, $\bu = A_j {\bu'}^{\sigma_j}$ for some $A_j\in Gl_d (k)$, and
$N_i = \la A_j^t \Gamma^t_i \hat{m}_i, {\bu'}^{\sigma_j} \ra$.  It is unlikely $A_j^t \Gamma^t_i ={\Gamma'}_b$ for some $b\in K^*$ where
$b\bu'={\Gamma'}_b \bu'$.

For a similar reason $N'_i$ is most likely not a $K$-global descent with respect to $\bu^{\sigma_j}$ and ${\bu'}^{\sigma_j}$ for $j=0,\ldots,d-1$. .

We have proved the following:
\begin{proposition}
\label{mixed-F}
Suppose $F(x,y)\in K[x,y]$ with $x=x_1,\ldots,x_n$ and $y=y_1,\ldots,y_n$.  For $i=0,\ldots,d-1$, we can efficiently construct $G(\hat{x},\hat{y})\in K[\hat{x},\hat{y}]$ such that $G(\hat{x},\hat{y}) = F^{\sigma_i} (\delta^{\sigma_i}\hat{x},{\delta'}^{\sigma_i} \hat{y})$ for all $(\hat{x},\hat{y})\in \hat{V}\times\hat{V'}$, and $G$ is of the form $\sum_i N_i N'_i$ with
$N_i = \la \Gamma^t_i \hat{m}_i, \bu\ra$ where $m_i$ is a monomial in $x$, $\Gamma_i$ is a $d$ by $d$ matrix with entries in $k$ and $N_i$ is not a $K$-global descent with respect to  $\bu^{\sigma_j}$ and ${\bu'}^{\sigma_j}$ for $j=0,\ldots,d-1$.  Similarly $N'_i=\la {\Gamma'}^t_i \hat{m'}_i, \bu'\ra$ where $m'_i$ is a monomial in $y$, ${\Gamma'}_i$ is a $d$ by $d$ matrix with entries in $k$ and $N'_i$ is not a $K$-global descent with respect to  $\bu^{\sigma_j}$ and ${\bu'}^{\sigma_j}$ for $j=0,\ldots,d-1$.
\end{proposition}

We say that the descent function $F^{\sigma_i}\circ (\delta^{\sigma_i},{\delta'}^{\sigma_i})$ in Proposition~\ref{mixed-F} is {\em properly specified} by $G$.  Note that when $G=\sum_i N_i N'_i$ is specified, $N_i$ is specified in the public basis $\bt$, that is,
$N_i = \sum_{j=0}^{d-1} f_{ij}\theta_j$ where $f_{ij}$ is a polynomial in $x_1,\ldots,x_n$ with coefficients in $k$ and $supp f_{ij}\subset supp \hat{m}_i$.
Similar observation applies to $N'_i$.

With specialization at $\beta\in \hat{V'}$, $G(\hat{x},\delta\beta) = \sum_i N_i N'_i (\delta\beta)$ takes the form
$\sum_j \la A^t_j \hat{m}_j, \bu\ra$ where $A_j$ is a heuristically random sum of matrices in $Gl_d (k)$, hence
$\la A^t_j \hat{m}_j, \bu\ra$ is unlikely a global $K$-global descent with respect to $\bu$.  Similar observation can  be made concerning specialization at $\alpha\in \hat{V}$.

\subsection{Blinding by Weil descent}
\label{blinding}
In this subsection we develop a method for blinding maps using Weil descent, to be employed later in our trilinear map construction.
Let $V\subset \bar{k}^n$ be an algebraic set defined as the zero set $Z(S)$ of a finite set $S$ of polynomials in $K[x_1,\ldots,x_n]$.
Let $\hat{V}\subset \bar{k}^{nd}$ be the descent of $V$, defined as the zero set $Z(\hat{S})$ where $\hat{S}$ contains all polynomials in $\hat{F}$ for every $F\in S$.

Suppose $m:V\times V\to V$ is a rational map defined over $K$.
Let $M$ be a $d\times d$ $(0,1)$-matrix such that each row has at most two nonzero entries, that is, entries with 1.
For row $i$, let $i_1$ and $i_2$ be such that  $0\le i_1\le i_2 \le d-1$ and $M(i,i_1) = M(i,i_2)=1$.

Let $\varphi:V^d \to V^d$ be a map determined by $M$ as follows.  Let $X=(X_i)_{i=0}^{d-1}$ with $X_i \in V(\bar{k})\subset \bar{k}^n$,
\[ \varphi (X) = (\varphi_i (X))_{i=0}^{d-1} \] where
\[\varphi_i (X) = m (X_{i_1}, X_{i_2})\]  for $i=0,\ldots, d-1$.

Let $\lambda: \hat{V}(K) \to V^d (K)$ be such that $\lambda = (\prod_{i=1}^{d-1} \sigma_{-i}) \circ \rho$.
Let $\hat{X}\in \hat{V}(K)\subset K^{nd}$.  Then, $\lambda (\hat{X}) = (\lambda_i (\hat{X}))_{i=0}^{d-1}$ where
$\lambda_i = \sigma_{-i}\circ \delta^{\sigma_i}$.

Let $\Psi:\hat{V}(K)\to \hat{V}(K)$ be such that $\Psi = \lambda^{-1}\circ \varphi\circ\lambda$.
We say that the map $\varphi$ is {\em blinded} by $\Psi$ with respect to $\bu$.

We have the following commutative diagram:
\[
\begin{array}{llll}
\hat{V}(K) & \stackrel{\rho}{\to} & \prod_{i} V^{\sigma_i} (K) \stackrel{\prod_i\sigma_{-i}}{\to}  & V(K)^d\\
\downarrow \Psi & &   & \downarrow \varphi\\
\hat{V}(K) & \stackrel{\rho}{\to} & \prod_{i} V^{\sigma_i} (K) \stackrel{\prod_i \sigma_{-i}}{\to}   & V(K)^d

\end{array}
\]

Let $A=(a_{ij})_{0\le i,j\le d-1}$ be a $d$ by $d$ matrix.  For $I\subset\{ 0,\ldots, d-1\}$, let $A^I = (a_{ij})_{0\le i \le d-1, j\in I}$, and $A_I = (a_{ij})_{i\in I, 0\le j \le d-1}$.

Suppose $A$ and $B$ are two $d$ by $d$ matrices.  If $I=\{ i\}$, then $A^I B_I=(c_{rs})_{0\le r,s\le d-1}$ with $c_{rs}=a_{ri}b_{is}$, the tensor product of the $i$-th column of $A$ and the $i$-th row of $B$.  In general, $A_I B_I =\sum_{i\in I} C(i)$ where $C(i)$ is the tensor product of the $i$-th column of $A$ and the $i$-th row of $B$.

Let $E=\{ (i-i_1\mod d, i-i_2\mod d : i=0,\ldots,d-1\}$.  For $(a,b)\in E$, let
$I_{a,b}=\{i: (i-i_1,i-i_2) = (a,b)\mod d \}$.

Let $\Omega_{a,b}=W^{I_{a,b}} \Gamma_{I_{a,b}}$, and $\Omega_{r,a,b}$ be the $r$-th row of $W^{I_{a,b}} \Gamma_{I_{a,b}}$.

We assume
\begin{itemize}
\item
$\bu$ is secret (so are $\Gamma$ and $W$),
\item
$\varphi$ is secret (so is $\{ (i,i_1,i_2): i=0,\ldots,d-1\}$),
\item
$\hat{m}$ is public.
\end{itemize}

\begin{proposition}
\label{specify-psi}
\[ \Psi_r (\hat{X}) = \sum_{(a,b)\in E} \la \hat{m} (\hat{X}^{q^a},\hat{X}^{q^b}), \Omega_{r,a,b}\ra\]
Consequently, $\Psi$ can be specified to the public by specifying $\hat{m}$, and making public $E$, and $\Omega_{a,b}$ for every $(a,b)\in E$.
\end{proposition}
\ \\{\bf Proof}
For $\hat{X}\in \hat{V}(K)$, $\Psi (\hat{X}) = (\Psi_i (\hat{X}))_{i=0}^{d-1}$ where
\begin{eqnarray*}
\Psi_r (\hat{X}) & = & \sum_{i=0}^{d-1} w_{ri} m^{\sigma_i} (\delta^{\sigma_i} \hat{X}^{\sigma_{i-i_1}}, \delta^{\sigma_i}\hat{X}^{\sigma_{i-i_1}}) \\
& = & \sum_{i=0}^{d-1} w_{ri} \la \hat{m} (\hat{X}^{\sigma_{i-i_1}}, \hat{X}^{\sigma_{i-i_2}}), \bu^{\sigma_i} \ra
\end{eqnarray*}
since
\[ m^{\sigma_i} (\delta^{\sigma_i} \hat{X}, \delta^{\sigma_i}\hat{Y}) = \la\hat{m}(\hat{X},\hat{Y}), \bu^{\sigma_i}\ra .\]

Hence

\[ \Psi_r (\hat{X}) = \sum_{(a,b)\in E} \la \hat{m} (\hat{X}^{q^a},\hat{X}^{q^b}), \Omega_{r,a,b}\ra\]
where
\[ \Omega_{r,a,b}=\sum_{i\in I_{a,b}} w_{ri} \bu^{\sigma_i}\]
which is the $r$-th row of $\Omega_{a,b} = W^{I_{a,b}} \Gamma_{I_{a,b}}$.
$\Box$

\ \\{\bf Some observations}
\begin{enumerate}
\item
If $| I_{a,b} | = 1$, then $\bu^{\sigma_i}$ can be determined up to constant factors (though $i$ is not known), then $U/u_0$ can be determined.  Therefore to keep $\bu$ secret, $| I_{a,b} |$ should be greater than.
\item
If $| I_{a,b} | = 2$ then for each of $O(d^2)$ possible choices of $I_{a,b}$ we are led to a system of $d^2$ quadratic equations in $4d$ unknown.  Similarly if $1 < | I_{a,b} | = O(1)$ then for each of $d^{O(1)}$ possible choices of $I_{a,b}$ we are led to a system of $d^2$ quadratic equations in $O(d)$ unknown. Solving such systems is difficult in general.
\item
If $|I_{a,b}|$ is big, say $| I_{a,b} | = o(d^c)$ for some positive constant $c < 1$, exhaustively trying all possible choices of $I_{a,b}$ is too costly.
\end{enumerate}

Suppose $d^{O(1)}$ many maps like $\varphi$ are blinded with respect to $\bu$.  We are led to the following:

\ \\{\bf Problem:}  A basis $\bu$ of $K$ over $k$ is hidden.  As before let $\Gamma$ be the matrix whose $i$-th row is $\bu^{\sigma_i}$ for $i=0,\ldots,d-1$.  Let $W=\Gamma^{-1}$.  A set of $d^{O(1)}$ matrixes is given, each of which is $W^I \Gamma_I$ for some secret $I\subset\{0,\ldots, d-1\}$ with $| I| = \Theta(d^c)$ for some positive constant $c < 1$.  Can $\bu$ be determined efficiently?
\subsection{Specifying maps on abelian varieties}
\label{specify-m}
A semi-algebraic set defined over $K$ in $\bar{k}^n$ is of the form $V(F_1,\ldots,F_m)- V(G_1,\ldots,G_r)$ where $F_i,G_j\in K[x_1,\ldots,x_n]$ for all $i,j$.
We may assume that an abelian variety $A$ can be described in terms of affine pieces.  As we will see in \S~\ref{jac} when we take $A$ to be the Jacobian variety of a hyperelliptic curve, we may assume $A (\bar{k}) = \cup_{i} V_{i}$ as a disjoint union, withe each $V_{i}$ an algebraic subset of  $\bar{k}^n$ for some $n$.  Moreover there is a unique $V_i$, say $i=0$, with $\dim V_0 =\dim A$. We call $V_0$ the {\em principal site} for $A$.

The addition morphism $m$ on $V_i\times V_j$ can be described in terms of a collection maps $m_{\alpha}: U_{\alpha} \to A$ where
$U_{\alpha}$ is a semi-algebraic subset of $V_i\times V_j$, and there is a unique $\alpha$ such that $U_{\alpha}$ is of the same dimension as $V_i\times V_j$, which we call the principal site for $m$ on $V_i\times V_j$.
The {\em principal site} for $m$ on $V_0\times V_0$ is the unique site of the same dimension as $A\times A$, and is called the {\em principal site} for $m$.

Similarly, the doubling morphism, sending $P\in A(\bar{k})$ to $2P$, has a principal site on $V_i$ for all $i$, and the principal site for the doubling morphism on $V_0$ is called the principal site for the morphism.

A point $\hat{P}\in\hat{A}(\bar{k})$ is said to be in a {\em pure site} of $\hat{A}$ if there is some $V_i$ such that  $\delta^{\sigma_j}\hat{P}\in V_i^{\sigma_j}$ for all $j$.  It is in a {\em pure site} for the doubling morphism if
if there is some $V_i$ such that for all $j$,
$\delta^{\sigma_j}\hat{P}$ is in the principal site of  $V_i^{\sigma_j}$ for $\sigma_j$-conjugate of the doubling morphism.

Suppose $\hat{P}_1,\hat{P}_2\in\hat{A}(\bar{k})$.  Then $(\hat{P}_1,\hat{P}_2)$ is in a pure site for $\hat{m}$ if there is some $V_i\times V_j$ such that $(\delta^{\sigma_r}\hat{P}_1,\delta^{\sigma_r}\hat{P}_2)$ is in the principal site of
$m^{\sigma_r}$ on $V_i^{\sigma_r}\times V_j^{\sigma_r}$ for all $r$. If $i=j=0$, then it is said to be in the principal site for $\hat{m}$.

Suppose  $\hat{P}_1,\hat{P}_2\in\hat{A}(\bar{k})$.

Suppose $(\hat{P}_1,\hat{P}_2)$ belongs to a pure site for $\hat{m}$.  Then there is
$m_{\alpha}: U_{\alpha}\to V$,
$(\delta^{\sigma_i}\hat{P}_1,\delta^{\sigma_i}\hat{P}_2)\in U_{\alpha}$ for all $i$, and $U_{\alpha}$ is the principal site for $m$ on some $V_j \times V_r$.  In this case, $\hat{m} (\hat{P}_1,\hat{P}_2)=\hat{m}_{\alpha} (\hat{P}_1,\hat{P}_2)$.

More generally, if $(\delta^{\sigma_i}\hat{P}_1,\delta^{\sigma_i}\hat{P}_2)\in U_{\alpha_i}$,  then $\hat{m} (\hat{P}_1,\hat{P}_2)
=\Gamma^{-1} v$ where $v=(v_i)_{i=0}^{d-1}$ and
\[ v_i = m_{\alpha_i}(\delta^{\sigma_i}\hat{P}_1,\delta^{\sigma_i}\hat{P}_2)=\la \hat{m}_{\alpha_i} (\hat{P}_1,\hat{P}_2),\bu^{\sigma_i} \ra \]

Then
\[
\begin{array}{lll}
\delta^{\sigma_i} \hat{m}(\hat{P}_1,\hat{P}_2) & = & \sum_{j=0}^{d-1} w_{ij} v_j = \sum_{j=0}^{d-1} w_{ij} \la\hat{m}_{\alpha_j}(\hat{P}_1,\hat{P}_2),\bu^{\sigma_j} \ra \\
& = & \sum_{\alpha}  \la \hat{m}_{\alpha} (\hat{P}_1,\hat{P}_2), \sum_{j,\alpha_j =\alpha} w_{ij}\bu^{\sigma_j} \ra \\
& = & \sum_{\alpha} \la \hat{m}_{\alpha} (\hat{P}_1,\hat{P}_2), (\Omega_{\alpha})_i \ra
\end{array}
\]
where $\Omega_{\alpha}=W^{I_{\alpha}} \Gamma_{I_{\alpha}}$ with $I_{\alpha} =\{ i : 0\le i\le d-1, \alpha_i =\alpha \}$.

If $(\hat{P}_1,\hat{P}_2)$ belongs to a pure site for $\hat{m}$, then $\alpha_i = \alpha$ for all $i$, for some $\alpha$, $I_{\alpha}=\{0,\ldots,d-1\}$,
$\Omega_{\alpha}$ is the identity matrix, and we get $\hat{m}(\hat{P}_1,\hat{P}_2)=\hat{m}_{\alpha} (\hat{P}_1,\hat{P}_2)$ as already discussed.  If $(\hat{P}_1,\hat{P}_2)$ belongs to a mixed site for $\hat{m}$, then
the partition of $\{0,\ldots,d-1\}$ into $I_{\alpha}$'s, together with $\Omega_{\alpha}$ and $m_{\alpha}$ specifies the mixed site containing $(\hat{P}_1,\hat{P}_2)$.

We note the difference between this situation and the situation that arises in \S~\ref{blinding} is that in this case $I_{\alpha}$ also needs to be made public, consequently $\Omega_{\alpha}$ reveals a linear relation among
$W^{(i)}\Gamma_{(i)}$, $i=0,\ldots,d-1$.  Therefore the number of specified mixed sites should be carefully limited
so that the publicized set of $(I_{\alpha},\Omega_{\alpha})$ yields a small number of relations. In our situation, it is enough to focus on the principal sites for $\hat{A}$, $\hat{m}$ and the doubling morphism, hence there is no need to publicize any mixed site.  In this case we may focus on the principal site $V=V_0$ of $A$, consider $m:V\times V\to V$ and the doubling map $V\to V$ as rational maps, and focus on their descent maps and functions over $\hat{V}$.

\section{A trapdoor discrete logarithm problem}
\label{trapdoor-dl}
We apply the blinding method of \S~\ref{blinding} to define a trapdoor discrete-logarithm problem.

As in \S~\ref{blinding}, let $V\subset \bar{k}^n$ be an algebraic set defined by a finite set of polynomials in $K[x_1,\ldots,x_n]$. In the current context we assume that $V$ describes an affine piece of an abelian variety $A$ defined over $K$.  We assume that $A[\ell]\subset V(K)$.

Suppose $m:V\times V\to V$ is a rational map defined over $K$ that describes the addition morphism of $A$ when restricted to $V$.

The descent $\hat{m}:\hat{V}\times\hat{V}\to\hat{V}$ is formed in secret using $\bu$, and properly specified to the public,  so that
the specification does not contain any global descent and the entries in the matrix $W=\rho^{-1}$ are blinded, following the methods in Proposition~\ref{specify-set}, Proposition~\ref{specify-poly} and Proposition~\ref{specify-rat}.

We consider $(0,1)$-matrices $M$ with the property that
there are exactly two nonzero entries $(i,i_1)$ and $(i,i_2)$ for row $i$, for $i=0,\ldots,d-1$.
Let $E_M=\{ (i-i_1\mod d, i-i_2\mod d : i=0,\ldots,d-1\}$.  For $(a,b)\in E_M$, let
$I_{M,a,b}=\{i: (i-i_1,i-i_2) = (a,b)\mod d \}$.

Choose a set of $N=O(d^2)$ such matrices $M_1,\ldots,M_N$ such that
\begin{enumerate}
\item
$| I_{M_i,a,b} | = \Theta(d^{\epsilon}) $ for all $(a,b)\in E_{M_i}$, for some positive constant $\epsilon < 1$,
\item
the identity matrix and $M_1,\ldots,M_N$ span $Mat_d (\F_{\ell})$.
\end{enumerate}

Let for $i=1,\ldots,N$, $\varphi_i = \varphi_{M_i}$ be the map determined by $M_i$ and let
$\Psi_i$ be the map on $\hat{V}(K)$ blinding $\varphi_i$, as described in \S~\ref{blinding}.
By Proposition~\ref{specify-psi} $\Psi_i$ can be specified to the public by specifying $\hat{m}$, and making public $E_{M_i}$, and $\Omega_{a,b}$ for every $(a,b)\in E_M$. The property that $| I_{M_i,a,b} | = \Theta(d^{\epsilon}) $ is to make sure that the blinding of $\varphi_i$ is strong so that $M_i$ is hidden.

Find $\alpha,\beta\in A(K)[\ell]$ such that $e_{\ell} (\alpha,\beta)\neq 1$.  Then $\alpha$ and $\beta$ are not in the same cyclic group.
Choose random $x_i,y_i\in\F_{\ell}$ such that and let $D_{\beta}\in\hat{A}[\ell]$ such that $D_{\beta}$ corresponds to $V=(x_i \alpha+ y_i \beta)_{i=0}^{d-1}\in A[\ell]^d$.  We impose the condition that for some $i,j$, $x_i\alpha+y_i\beta
\neq x_j\alpha+y_j\beta$.  This is to make sure that $D_{\beta}$ is not a descent point, that is, there is no $\gamma\in A[\ell]$ such that  $\rho (D_{\beta})=(\gamma^{\sigma_i})_{i=0}^{d-1}$.

Let $M_0$ be the identity matrix and correspondingly $\varphi_0 =1$.

Let $\Lambda$ be the non-commutative $\F_{\ell}$-algebra generated by $N$ variables $z_1,\ldots,z_N$.

Let $\lambda:\Lambda\to \End \hat{A}(K)[\ell]$ be the algebra morphism defined by $\lambda z_i = \Psi_i$ for $i=1,\ldots,N$.

Let $\omega:\Lambda\to Mat_d (\F_{\ell})$ be the algebra morphism defined by $\omega(z_i)=M_i$ for $i=1,\ldots,N$.

\subsection{Forming quadratic relations}
To form a quadratic relation we choose random $a_{ij}$ and compute the matrix
$M=\sum_{1\le i,j\le N} a_{ij} M_i M_j$.  Then solve for $b_i$ such that
$M=b_0 + \sum_{1\le i\le N} b_i M_i$.  Hence a polynomial $R=\sum_{1\le i,j\le N} a_{ij} z_i z_j -\sum_{i=0}^N b_i z_i$ is
determined such that $R(M_1,\ldots,M_N) = 0$.  Let $supp_i f$ denote the subset of $supp f$ consisting of degree $i$ monomials.  Then $supp_1 R$ likely contains most of $z_1,\ldots,z_N$.

Form a set of $N_1=O(N)$ relations as above.  For simplicity suppose $N_1 = N$ and let $\cR=\{R_1,\ldots, R_N\}$ be the set of relations which are formed.

\subsection{The discrete logarithm problem}

Let $J$ be the two-sided ideal of $\Lambda$ generated by $\cR$.

For $i > 0$,
let $J_i$ be the submodule of $J$ consisting of elements of degree less than or equal to $i$.

Let $G=U_1/U\cong \Z/\ell\Z$ where $U=J_N$ and $U_1 = \F_{\ell}+U$.
The discrete logarithm problem on $G$ is formally the problem of computing the map $G\to\Z/\ell Z$ sending $a+U\in G$ to $a$ for $a\in\F_{\ell}$.

We specify
the discrete logarithm problem on $G$ as follows.
\begin{enumerate}
\item
The set $\cR$ is made public, and $J$ is specified as the two sided ideal of $\Lambda$ generated by $\cR$.
\item
The group $G$ is defined as $U_1/U$, where $U=J_N$, $U_1=\F_{\ell}+U$.
For $a\in\F_{\ell}$, $a+U\in G$ is encoded by a sparse representative in $a+U$.  More precisely, to encode $a$, one follows the procedure described in \S~\ref{sparse} to construct a sparse element $f\in U$ with $|supp f|=O(N^2)$.  Let $g=\sum_i a_i m_i =f+a$, where $m_i$ are monomials of degree no greater than $N$.   Then $g$ is an encoding of $a$.

\item
The discrete logarithm problem on $G$ is: Given a sparse $g\in U_1$, to determine $a\in\F_{\ell}$ such that $g\in a+U$.
\end{enumerate}

The morphism $\omega$ is a trapdoor map since $\omega g = a I$ where $I$ is the identity matrix.

The discrete logarithm as specified above is the generic version that does not involve the abelian variety $A$, and the maps $\Psi_i$ on $\hat{A}[\ell]$.  In this generic version the first condition in forming $M_i$ is not needed.

When the maps $\Psi_i$ are specified together with $D_{\beta}\in\hat{A}[\ell]$, public identity testing for $G$ is made possible:  for $g\in U_1$, $g\equiv 0 \mod U$ if and only if $\lambda(g) (D_{\beta})=0$.  We call this version {\em trapdoor discrete logarithm on G with public identity testing}.

\subsection{Constructing random sparse elements in $J_N$}
\label{sparse}
We call an element $f\in\Lambda$ $s$-sparse if $| supp f | \le s$.
We describe a method to construct an $O(N^2)$-sparse $f\in J_N$ randomly with $f=\sum_{i=1}^{N-1} f_i$ so that
\begin{enumerate}
\item
$f_i\in J_{i+1}$ for $i=1,\ldots, N-1$,
\item
$supp f_i$ consists of monomials of degree $i-1, i, i+1$ for $i=1,\ldots, N-1$,
\item
$| supp f_i \cap supp f_{i+1} | \ge N^c$ for some constant $0 < c <1$, for $i=1,\ldots, N-2$.
\end{enumerate}

To construct $f$ the first step is to form $f_1$ as a random $\F_{\ell}$ linear combination of $R_1,\ldots,R_N$.
Then proceed inductively to form $f_i$ for $i=2,\ldots,N-1$.
Suppose $f_{i-1}$ has been determined.  To form $f_i$ we do the following.
\begin{enumerate}
\item
Form $R^{(i)}_j = \sum_k r_{ijk} R_k$ with random $r_{ijk}\in\F_{\ell}$ for $1\le j,k \le N$.
\item
For $j=1,\ldots,N$, choose two random monomials $m_{1j}$ and $m_{2j}$ such that $\deg m_{1j}m_{2j} = i-1$, and set $f^{(i)}_{j}=m_{1j} R^{(i)}_j m_{2j}$.
\item
Randomly choose $N^c$ terms in $f_{i-1}$ of degree $i$.  For each chosen term $t$, find some $R^{(i)}_j$ such that some variable $z_k\in supp_1 R^{(i)}_j$ appears in $t$.  Write $t=a m_1 z_k m_2$ where $a\in\F_{\ell}$ and $m_1,m_2$ are monomials with $\deg m_1m_2 = i-1$.  Set $g_t = m_1 R^{(i)}_j m_2$.  Form
$G^{(i)}=\sum_t r_t g_t$ with random $r_t\in\F_{\ell}$ and $t$ ranges over the $N^c$ chosen terms.
\item
Set $f_i = \sum_{j=1}^N a_j f^{(i)}_j + G^{(i)}$ with randomly chosen $a_j\in\F_{\ell}$ for $j=1,\ldots,N$.
\end{enumerate}

Note that $|supp f | = O(N^2)$.

\section{Trilinear maps involving Weil descent}
\label{tri-weil}
\subsection{Constructing the trilinear map}
To construct a trilinear map, we take an abelian  variety $A$ of dimension $g$ defined over a finite field $K$ of extension degree $d$ over a finite field $k$,
and consider the descent $\hat{A}$ of $A$ with respect to a random secret basis $\bu$ of $K$ over $k$.
The descent $\hat{A}$ and $\hat{m}$ are specified to the public in such a way that the specification does not contain any global descent.
For simplicity assume  $\log\ell$, $d$ and $\log | k |$ are linear in the security parameter $n$, whereas $g=O (1)$.

The trilinear map will take the form $G_1\times G_2\times G_3\to \mu_{\ell}\subset K$ where $G_1$ is a cyclic group generated by a point $D_{\alpha}\in \hat{A}(K)[\ell]$,  $G_2$ is a cyclic group generated by a point $D_{\beta}\in \hat{A}(K)[\ell]$, and $G_3$ is a cyclic group with a trapdoor as discussed in \S~\ref{trapdoor-dl}.

As in \S~\ref{trapdoor-dl}, we choose a set of $N=O(d^2)$ many $(0,1)$-matrices $M_1,\ldots,M_N$ that span $Mat_d (\F_{\ell})$, so that each $M=M_i$ has the following properties:
\begin{enumerate}
\item
There are exactly two nonzero entries $(i,i_1)$ and $(i,i_2)$ for row $i$, for $i=0,\ldots,d-1$.
\item
Let $E_M=\{ (i-i_1\mod d, i-i_2\mod d : i=0,\ldots,d-1\}$.  For $(a,b)\in E_M$, let
$I_{M,a,b}=\{i: (i-i_1,i-i_2) = (a,b)\mod d \}$.
Then $| I_{M_i,a,b} | = \Theta(d^{\epsilon}) $ for all $(a,b)\in E_{M_i}$, for some positive constant $\epsilon < 1$,
\end{enumerate}

Let $\Lambda$ be the non-commutative $\F_{\ell}$-algebra generated by $N$ variables $z_1,\ldots,z_N$.

Let $\lambda:\Lambda\to \End \hat{A}(K)[\ell]$ be the algebra morphism defined by $\lambda z_i = \Psi_i$ for $i=1,\ldots,N$.

We have the following commutative diagram:

\[
\begin{array}{llll}
\hat{A}[\ell] & \stackrel{\rho}{\to} & \prod_{i} A^{\sigma_i} [\ell]\stackrel{\prod_i \sigma_{-i}}{\to}  & A[\ell]^d\\
\downarrow \Psi_i & &   & \downarrow \varphi_{M_i}\\
\hat{A}[\ell] & \stackrel{\rho}{\to} & \prod_{i} A^{\sigma_i} [\ell] \stackrel{\prod_i \sigma_{-i}}{\to} & A[\ell]^d

\end{array}
\]

Let $e_{\ell}: A[\ell]\times A[\ell]\to \mu_{\ell}$ be a nondegenerate bilinear pairing.
On $\hat{A}[\ell]$ we define for $\hat{D}_1,\hat{D}_2\in\hat{A} [\ell]$,
\[ \hat{e} (\hat{D}_1,\hat{D}_2) = \prod_{i=0}^{d-1} e_i ( \delta^{\sigma_i} \hat{D}_1,\delta^{\sigma_i}\hat{D}_2)\]
where $e_i= e^{\sigma_i}_{\ell}$.  Note that $\hat{e}$ is the blinded version of the pairing $\prod_{i=0}^{d-1} e_i$
on $\prod_{i=0}^{d-1} A^{\sigma_i}[\ell]$.

Find $\alpha,\beta\in A(K)[\ell]$ such that $e_{\ell} (\alpha,\beta)\neq 1$.  Then $\alpha$ and $\beta$ are not in the same cyclic group.
Choose random $x_i,y_i\in\F_{\ell}$ such that and let $D_{\alpha}\in\hat{A}[\ell]$ such that $D_{\alpha}$ corresponds to $V=(x_i \alpha+ y_i \beta)_{i=0}^{d-1}\in A[\ell]^d$.  We impose the condition that for some $i,j$, $x_i\alpha+y_i\beta
\neq x_j\alpha+y_j\beta$.  This is to make sure that $D_{\alpha}$ is not a descent point, that is, there is no $\gamma\in A[\ell]$ such that  $\rho (D_{\alpha})=(\gamma^{\sigma_i})_{i=0}^{d-1}$.

Similarly, choose random $x'_i,y'_i\in\F_{\ell}$ such that and let $D_{\beta}\in\hat{A}[\ell]$ such that $D_{\beta}$ corresponds to $V=(x'_i \alpha+ y'_i \beta)_{i=0}^{d-1}\in A[\ell]^d$.  We impose the condition that for some $i,j$, $x'_i\alpha+y'_i\beta
\neq x'_j\alpha+y'_j\beta$, so that $D_{\beta}$ is not a descent point.

Furthermore, $x_i,y_i,x'_i,y'_i$ are chosen so that
\[\hat{e} (D_{\alpha}, D_{\beta})=\prod_i (e_{\ell}( (x_i \alpha + y_i \beta ), (x'_i \alpha + y'_i \beta ) ))^{\sigma_i} \neq 1.\]

Form, as in \S~\ref{trapdoor-dl}, a set $\cR$ of $N_1=O(N)$ dense quadratic relations on $M_1,\ldots,M_N$.  For simplicity suppose $N_1 = N$ and let $\cR=\{R_1,\ldots, R_N\}$ be the set of relations which are formed.
Let $J$ be the two-sided ideal of $\Lambda$ generated by $\cR$.

Let $G_1$ be the group generated by $D_{\alpha}$.
Let $G_2$ be the group generated by $D_{\beta}$.
Let $G_3=U_1/U\cong \Z/\ell\Z$ where $U=J_N$ and $U_1 = \F_{\ell}+U$.

The trilinear map $G_1\times G_2 \times G_3 \to \mu_{\ell}$ sends
$(xD_{\alpha},yD_{\beta},z+U)$ to $\zeta^{xyz}$ where $\zeta=\hat{e}(D_{\alpha},D_{\beta})$.
Suppose $z+U$ is represented by some sparse $\gamma\in z+U$.  Then
\[ \hat{e}(xD_{\alpha},\lambda(\gamma)(yD_{\beta}))=\hat{e} (xD_{\alpha},zyD_{\beta})=\zeta^{xyz}.\]

The sparsity constraint is to make sure that the map $\gamma$ can be efficiently executed,so that the trilinear map can be efficiently computed, assuming the pairing is efficiently computable.

We note that if the two secret descent bases were identical then the published pairing $\hat{e}$ together with some $\Psi_i$ can be used to induce self pairing on $G_1$.  Namely if $\hat{e} (D_{\alpha}, \Psi_i (D_{\alpha}))\neq 1$, then we have an efficiently computable pairing $G_1\times G_1 \to \mu_{\ell}$, hence $G_1$ would not satisfy DDH assumption.  Similar observation applies to $G_2$.  As for $G_3$, neither the pairing $\hat{e}$ nor the trilinear map naturally induce a self pairing on the group.

In order for the cyclic groups $G_1$ and $G_3$ to satisfy the DDH assumption, we can construct the two groups on two descent $\hat{A}$ and $\hat{A}'$ of $A$ with respect two secret bases. Then the pairing $\hat{e}: \hat{A}[\ell]\times\hat{A}'[\ell]\to \mu_{\ell}$ cannot be used to define a self pairing on $G_1$ or $G_2$ directly.

In this setting, we
fix a public basis $\theta_1,\ldots,\theta_d$ of $K/k$, a private basis $u_1,\ldots,u_d$ of $K/k$, and another private basis $u'_1,\ldots,u'_d$ of $K/k$.

Let $\delta$ denote the basic descent map $\bar{k}^d \to \bar{k}$ with respect to $u_1,\ldots,u_d$, and $\rho$ the bijective linear map $\bar{k}^d \to \bar{k}^d$ determined by $\delta$.

Let $\delta'$ denote the basic descent map $\bar{k}^d \to \bar{k}$ with respect to $u'_1,\ldots,u'_d$, and $\rho'$ the bijective linear map $\bar{k}^d \to \bar{k}^d$ determined by $\delta'$.

Let $\hat{A}$ denote the descent of $A$ with respect to the basis $u_1,\ldots,u_d$.

Let $\hat{A}'$ denote the descent of $A$ with respect to the basis $u'_1,\ldots,u'_d$.

Then $\hat{e}: \hat{A}'[\ell]\times \hat{A}[\ell]$ is defined such that for $D_1\in\hat{A}'[\ell]$ , $D_2\in \hat{A}[\ell]$,
\[ \hat{e} (D_1,D_2) = \prod_{0\le i\le d-1} e_{i} (\delta'^{\sigma^i}(D_1), \delta^{\sigma^i} (D_2)).\]

We publish the following
\begin{enumerate}
\item  $D'_{\alpha}$ and $D_{\beta}$ where $D'_{\alpha}$ is the image of $D_{\alpha}$ under the natural isomorphism
between $\hat{A}$ and $\hat{A}'$ determined by $\rho'^{-1}\rho$,
\item the program for computing the descent $\hat{m}$ of the addition $m$ on $\hat{A}$,
the program for computing the descent $\hat{m}'$ of the addition $m$ on $\hat{A}'$,
\item  the programs for computing $\Psi_i$, $i=1,\ldots,N$,
\item  the set $\cR$ of relations
\end{enumerate}

We also need to specify $\hat{e}$ such that it is efficiently computable in the public while $\bu$ and $\bu'$ remain secret.
We will show how this can be done when $A$ is the Jacobian variety of a hyperelliptic curve in the next section.

\section{Jacobian varieties of hyperelliptic curves}
\label{jac}
We consider the Jacobian variety $J=J_C$ of a hyperelliptic curve $C$ of genus $g$ with an affine model $y^2 = f(x)$ where $f\in K[x]$ of degree $2g+1$ where $g >1$.  Again let $d = [K:k]$, and for simplicity assume  $\log\ell$, $d$ and $\log | k |$ are linear in the security parameter $n$, whereas $g=O (1)$.  All computations described below will take time polynomially bounded in $\log\ell$, $d$,
$\log | k |$,
and $g^{O(g)}$, hence  polynomially bounded in $n$.

We follow \cite{C} and consider the birational model for representing points of $J$ by reduced divisors on $C$.  Following \cite{C}, a {\em semireduced} divisor is of the form $\sum_{i=1}^r P_i -r \infty$, where if $P_i = (x_i,y_i)$ then $P_j \neq (x_i,-y_i)$ for $j\neq i$.  A semireduced divisor $D$ can be uniquely represented by a pair of polynomials $(a,b)$
such that $a(x)=\prod_{i=1}^r (x - x_i)$, $\deg (b) < \deg (a)$ , and $b^2 \equiv f \mod a$. We write $D=\Div (a,b)$. The divisor $D$ is $K$-rational if  $a,b\in K[x]$.  A reduced divisor is a semireduced divisor $D$ with $r\le g$, represented by a pair of polynomials $(a,b)$ where $\deg b < \deg a \le g$ and $a$ is monic.

To describe the sites of the Jacobian variety, let us consider briefly polynomial division.  Let $f$ and $g$ be polynomials of degrees $n$ and $m$ respectively.   Then $f= qg+r$ where $\deg q = n-m$ and $\deg r \le m-1$.  Let $(f_i)_{i=0}^n$, $(g_i)_{i=0}^m$, $(q_i)_{i=0}^{n-m}$ and $(r_i)_{i=0}^{m-1}$ be the coefficient vectors of $f,g,q,r$ respectively.  Then $q_{n-m-i}$ can be expressed as a rational function in $f_i$'s and $g_i$'s of degree $i+1$, for $i=0,\ldots, n-m$; and $r_i$ can be expressed as a rational function of degree $n-m+2$ for $i=0,\ldots, m-1$.  When $g$ is monic then $q_{n-m-i}$ can be expressed as a polynomial in $f_i$'s and $g_i$'s of degree $i+1$, for $i=0,\ldots, n-m$; and $r_i$ can be expressed as a polynomial of degree $n-m+2$ for $i=0,\ldots, m-1$.

A point on $J$ is represented by a reduced divisor $\Div (a,b)$ where $a$ is monic, $\deg a \le g$ and $\deg b \le \deg a -1$, satisfying $f\equiv b^2 \mod a$.  The last condition can be expressed by demanding the remainder of the division of $f-b^2$ by $a$ to be 0.  From the discussion above this translates into $\deg a$ polynomial conditions of degree $O(g)$, namely by setting the $\deg a$ many remainder polynomials to zero.  We have $g+1$ disjoint affine pieces $V_i$, $i=0,\ldots,g$, where $V_i$ corresponds to the case where $\deg a=g-i$.  Each piece is an algebraic subset of $\bar{k}^{2g+1}$.
A $K$-rational point of $J$ corresponds to a $K$-rational pair $(a,b)$, which can be naturally identified with a $K$-rational point in $K^{2g+1}$.
The principal site of $J$ is $V_0$, corresponding to the case $\deg a = g$.

The addition law can be described in terms of two algorithms: {\em composition} of semireduced divisors and {\em reduction} of a semireduced divisor to a reduced divisor \cite{C}.

Suppose $D_1 =\Div (a_1, b_1)$ and $D_2 = \Div (a_2, b_2)$ are two semireduced divisors.  Then $D_1 + D_2 = D + ( h )$ where $D = \Div (a,b)$ is semireduced and $h(x)$ is a function, and $a,b$ and $h$ can be computed by a
{\em composition} algorithm.  We have
\[ h = gcd (a_1, a_2,b_1+b_2 ) = h_1 a_1 + h_2 a_2 + h_3 (b_1+b_2)\]
where $h_1$, $h_2$ and $h_3$ are polynomials and $h$ is monic.

\[ a = \frac{a_1 a_2}{h^2} \]

\[ b = \frac{h_1 a_1 b_2 + h_2 a_2 b_1 + h_3 (b_1 b_2 +f)}{h} \mod a \]

Suppose $D=\Div (a,b)$ is a semireduced divisor with $\deg a > g$.  Then a {\em reduction} when applied to $D$ results in a smaller semi-reduced divisor $E=\Div (a',b')$ where
\[ a' = \frac{f-b^2}{a}\]
\[b'= -b \mod a',\]
and $D = E + (h')$
with $h'=\frac{y-b}{a'}$.  We have $\deg a' \le \deg a -2$.

If $D_1$ and $D_2$ are two reduced divisors then after a composition we get a semireduced divisor of degree at most $2g$.
So in $O(g)$ iterations of reductions we eventually obtained a reduced divisor $D_3$ and a function $h$ so that
$D_1 + D_2 = D_3 + (h)$.   We call this computation {\em addition}:  on input reduced divisors $D_1=\Div (a_1,b_1)$ and $D_2 = \Div (a_2,b_2)$, a reduced divisor $D_3 = \Div (a_3,b_3)$ together with a function $h$ are constructed, so that $D_1 + D_2 = D_3 + (h)$.

Note that the function $h$ is of the form $h_1 h_2$ where $h_1 (x)$ is a polynomial monic of degree less than $2g$ resulting from the composition step, and
$h_2$ is the product of $O(g)$ functions of the form $\frac{y-\beta(x)}{a' (x)}$, each resulting from a reduction step, where the degrees of $\beta (x)$ and $a'(x)$ are less than $2g$.

We define the degree of a rational function $f/g$, where $f$ and $g$ are polynomials, to be the maximum of $\deg f$ and $\deg g$.

We observe that the basic operations in composition and reduction are polynomial addition, multiplication and division (to obtain quotient and remainder).
The addition of two reduced divisors involves $O(g)$ polynomial divisions.  Each division leads to $O(g)$ branches of computation depending on the degree of the remainder.  The degrees of the coefficients of quotient and remainder polynomials as polynomials in the coefficients of $a_1$, $b_1$, $a_2$ and $b_2$ increase by a factor of $O(g)$ with each division.  From a routine analysis we see that the map $m$ on $V_i \times V_j$ can be divided into
$g^{O(g)}$ sites.  Each case is a rational map defined by $O(g)$ functions of degree $g^{O(g)}$ in the coefficients of $a_1,b_1,a_2,b_2$, and the semi-algebraic set for the site is defined by $g^{O(1)}$ polynomials of degree $g^{O(g)}$
in $a_1$, $b_1$, $a_2$ and $b_2$.

For an unknown reduced divisor $D=\Div (a,b)$ we let $x_D$ denote the list of variables representing the coefficients of $a$ and $b$.
From the addition algorithm and the analysis above, we see that the coefficients of $h_1 (x)$ and each $a'(x)$ and $\beta(x)$ are rational functions of degree $g^{O(g)}$ in $x_{D_1}$ and $x_{D_2}$.

 At the principal site of $m$ on $V_i\times V_j$, $h_1 = 1$, hence $h$  is the product of $O(g)$ functions of the form $\frac{y-\beta(x)}{a' (x)}$, each resulting from a reduction step, where the degrees of $\beta (x)$ and $a'(x)$ are less than $2g$,  and their coefficients are rational functions of degree $g^{O(g)}$ in $x_{D_1}$ and $x_{D_2}$.

Similarly at the principal site of the doubling map $2$ on $V_i$, if we write $2D = D' + (h)$ where $D$ is a reduced divisor at the site and $D'$ is the resulting reduced divisor. Then $h$  is the product of $O(g)$ functions of the form $\frac{y-\beta(x)}{a' (x)}$, where the degrees of $\beta (x)$ and $a'(x)$ are less than $2g$, with coefficients being rational functions of degree $g^{O(g)}$ in $x_{D}$.

Summarizing our discussion so far, we have the following.

\begin{proposition}
\label{addition}
\begin{enumerate}
\item
The addition of reduced divisors at a site $V_i\times V_j$, and similarly the doubling map at a site $V_i$, can be divided into
$g^{O(g)}$ cases.  Each case is a rational map defined by $O(g)$ functions of degree $g^{O(g)}$ on an algebraic set, and the algebraic set is defined by $g^{O(1)}$ polynomials of degree $g^{O(g)}$.

\item
If we write $D_1 +D_2 = D_3 + (h)$ where $D_i$ are reduced divisors for $i=1,2,3$, $(D_1,D_2)$ belongs to a site of $m$ on $V_i\times V_j$, and $h$ is a function, then
$h = h_1 h_2$, where $h_1 (x)$ is a polynomial monic of degree less than $2g$ resulting from the composition step, and
$h_2$ is the product of $O(g)$ functions of the form $\frac{y-\beta(x)}{a' (x)}$, each resulting from a reduction step. The degrees of $\beta (x)$ and $a'(x)$ are less than $2g$, and the coefficients of $h_1 (x)$ and each $a'(x)$ and $\beta(x)$ are rational functions of degree $g^{O(g)}$ in $x_{D_1}$ and $x_{D_2}$.
Moreover,
at the principal site of $m$ on $V_i\times V_j$, $h_1 = 1$.
\item
Similarly at the principal site of the doubling map $2$ on $V_i$, if we write $2D = D' + (h)$ where $D$ is a reduced divisor at the site and $D'$ is the resulting reduced divisor. Then $h$  is the product of $O(g)$ functions of the form $\frac{y-\beta(x)}{a' (x)}$, where the degrees of $\beta (x)$ and $a'(x)$ are less than $2g$, with coefficients being rational functions of degree $g^{O(g)}$ in $x_{D}$.
\end{enumerate}
\end{proposition}

In pairing computation we will need to evaluate the function $h$ on reduced divisors.  To this end it is sufficient to consider functions that are either polynomials in $x$, or of the form $y-\beta(x)$ where $\beta(x)$ is a polynomial in $x$.

Let $\nuinf$ denote the valuation on the function field of $C$ at infinity. Then $\nuinf(x)=-2$ and $\nuinf(y)=-(2g+1)$,
and $x^g y^{-1}$ is a local uniformizing parameter for $\nuinf$.

For functions $f$ and $g$ we write $f\siminf g$ if $\frac{f}{g} (\infty)=1$.

For $f\in K[x]$ let $f_{\infty}$ denote the leading coefficient of $f$.  Then $\nuinf (f) = -2\deg f$ and
$f \siminf f_{\infty} x^{\deg f}$.

We assume that the hyperelliptic curve is given by an equation $y^2 - f(x)$ where $\deg f=2g+1$ and $f$ is monic.
In this case if $\nuinf (x^a y^b)=0$ then $x^a y^b  (\infty)=1$.    This is because $\nuinf (x)$ is even and $\nuinf (y)$ is odd, so $b$ must be even.  Put $b=2c$.  Then $\nuinf (x^a y^b)=0$ implies $a+c(2g+1)=0$.  We have
$$x^a y^b = x^a y^{2} = x^a f^c \siminf  x^{a+c(2g+1)} =1.$$

Consider the function $y-b$ where $b\in K[x]$.  If $\deg b \le g$ then
$\nuinf (y-b) = \nuinf (y)= -(2g+1)$, and $\nuinf (y^{-1} b) > 0$.  We have
$\frac{y-b}{y} (\infty) = (1 - y^{-1} b ) (\infty) = 1$, so $y\siminf y-b$.

If $\deg b > g$ then $\nuinf (b^{-1} y ) > 0$.  We have $\frac{y-b}{b} (\infty) = (b^{-1} y -1 )(\infty) =-1$, so
$y-b\siminf -b$.

Suppose a function $h$ is of the form $h = \frac{h_1 (x)}{h_2 (x)} \prod_i y-\beta_i (x)$, where $h_1$, $h_2$ and $\beta_i$ are polynomials in $x$.  Then we have
\[ h\siminf \frac{(h_1)_{\infty}}{(h_2)_{\infty}} \prod_{i,\deg\beta_i > g} (-\beta_i)_{\infty} x^a y^b\]
where $a=\deg h_1 -\deg h_2 +\sum_{i,\deg \beta_i > g} \deg\beta_i$ and
$b$ is the number of $i$ with $\deg\beta_i \le g$.

We have the following.

\begin{lemma}
\label{h-inf}
\begin{enumerate}
\item
We assume that the hyperelliptic curve is given by an equation $y^2 - f(x)$ where $\deg f=2g+1$ and $f$ is monic.
In this case if $\nuinf (x^a y^b)=0$ then $x^a y^b  (\infty)=1$.
\item
Suppose a function $h$ is of the form $h = \frac{h_1 (x)}{h_2 (x)} \prod_i y-\beta_i (x)$, where $h_1$, $h_2$ and $\beta_i$ are polynomials in $x$.  Let $h_{\infty} = \frac{(h_1)_{\infty}}{(h_2)_{\infty}} \prod_{i,\deg\beta_i > g} (-\beta_i)_{\infty}$  Then $h\siminf h_{\infty} x^a y^b $,
where $a=\deg h_1 -\deg h_2 +\sum_{i,\deg \beta_i > g} \deg\beta_i$ and
$b$ is the number of $i$ with $\deg\beta_i \le g$.
\end{enumerate}
\end{lemma}

Consider now the evaluation of $h$, which is either a polynomial in $x$ or of the form $y-\beta(x)$ where $\beta(x)$ is a polynomial in $x$, at the affine part of a reduced divisor.

Let
\[ D=\Div (a',b') = \sum_i P_i - r\infty\] be a reduced divisor.
Then $y(P_i) = b' (P_i)$, so
\[ (y-\beta) (\sum_i P_i ) = (b' -\beta) (\sum_i P_i ) =\prod_i (b'-\beta) (\alpha_i)\]
where $a' (x) = \prod_i (x-\alpha_i)$.

Let $\Phi(x) =\sum_{i=0}^{2g-1} t_i x^i\in K[x,t_0,\ldots,t_{2g-1}]$.
We can construct by the fundamental theorem of symmetric polynomials a polynomial $S (\btt, \bz)$ where $\btt = t_0,\ldots,t_{2g-1}$ and $\bz= z_1,\ldots,z_g$, such that
\[S(\btt, s_1(\bz),\ldots,s_g(\bz)) = \prod_{i=1}^g \Phi(z_i)\]
where $s_i(\bz)$ is the $i$-th symmetric expression in $z_1,\ldots,z_g$ ($s_1 (\bz) = z_1 + \ldots+z_g$ for example).
The polynomial $S$ has degree $O(g)$ in $\btt$ and degree $O(g)$ in $\bz$.

Let $f=\sum_{i=0}^m a_i x^i \in K[x]$ of degree $m < 2g$.   Denote by
$c(f) = (a_0,\ldots,a_m,0,\ldots,0)$ the $(2g)$-vector consisting of the coefficients of $f$ padded with 0's if necessary.

Let $\rho (x)\in K[x]$ of degree $r \le g$ and monic.  Let $\gamma_1,\ldots,\gamma_r$ be the roots of $\rho$ and let  $\gamma = \gamma_1,\ldots,\gamma_r$.
Then
\[ \rho (x) = \prod_{i=1}^r (x-\gamma_i )=x^r+ \sum_{i=1}^r (-1)^i s_i (\gamma) x^{r-i}.\]

Let $s(\rho) = (s_1 (\gamma),\ldots,s_r (\gamma),0,\ldots,0)$, the $g$-vector consisting of $s_i (\gamma)$ and padded with 0 if necessary.

We have
\[ S(c(f),s(\rho))= \prod_{i=1}^r f (\gamma_i).\]

Therefore if $D=\Div (a,b)$ is a reduced divisor then $D= D^{+} - r\infty$ for some $r\le g$, then
\[ f(D^+) = S( c(f), s(a)).\]

For function $y-\beta(x)$ where $\deg \beta <2g$, then $y-\beta (D^+) = b-\beta (D^+)$.  Therefore
\[ (y-\beta) (D^{+} ) = S (c(b-\beta), s(a)).\]

We have proved the following.

\begin{lemma}
\label{hD+}
\begin{enumerate}
\item
Let $f=\sum_{i=0}^m a_i x^i \in K[x]$ of degree $m < 2g$.   Denote by
$c(f) = (a_0,\ldots,a_m,0,\ldots,0)$ the $(2g)$-vector consisting of the coefficients of $f$ padded with 0's if necessary.  If $D=\Div (a,b)$ is a reduced divisor, write $D= D^{+} - r\infty$ for some $r\le g$, then
$f(D^+) = S( c(f), s(a))$.
\item  For function $y-\beta(x)$, then $y-\beta (D^+) = b-\beta (D^+)$.  Therefore
\[ (y-\beta) (D^{+} ) = S (c(b-\beta), s(a)).\]

\end{enumerate}
\end{lemma}

Let $D_1=\Div (a_1,b_1)$ be a reduced divisor. Then $2D_1 = D' + (h)$ where $D'$ is a reduced divisor and $h$ is a function.
By Proposition~\ref{addition} we know that $h$ is of the form
${h_1}{h_2}$ where $h_1 \in K[x]$ is of degree less than $2g$ and
$h_2 = \prod_i \frac{y - \beta_i (x)}{a'_i (x)} $ where $\beta_i (x)$ and $a'_i (x)$ are polynomials in $x$, and  $\deg \beta_i$ and the number of $i$ are both less than $2g$.  Put $h_3 (x)=\prod_i a'_i (x)$.  Then $\deg h_3 (x)=O(g^2)$.

Let $h^+ (x_D)$ denote the function $h$ as it applies to evaluate the positive part $D^+$ of a reduced divisor $D=\Div (a,b)$.  Then by Lemma~\ref{hD+},

\[h^+ (x_D)= h(D^+) = \frac{S(c(h_1), s(a))}{S(c(h_3), s(a))}  {\prod_i S(c(b-\beta_i), s(a))}.\]

At each site of the doubling map, we have, by Proposition~\ref{addition},
\[
\begin{array}{lll}
h_1 (x) & = & \sum_{i=0}^{2g-1} \lambda_i (x_{D_1}) x^i\\
h_3 (x) & = & \sum_{i=0}^{d} \lambda'_i (x_{D_1}) x^i\\
\beta_i (x) & = & \sum_{j=0}^{2g-1} \mu_{ij}(x_{D_1}) x^j
\end{array}
\] where $d=O(g^2)$, $\lambda_i$, $\lambda'_i$ and $\mu_{ij}$ are rational functions of degree $g^{O(g)}$ in $x_{D_1}$.

Since $h_1$ is monic, $(h_1)_{\infty}=1$, and we have
\[ h_{\infty} = \frac{(h_1)_{\infty}}{(h_3)_{\infty}} \prod_{i,\deg\beta_i > g} (-\beta_i)_{\infty}
=\frac{\prod_{i,\deg\beta_i > g} (-\beta_i)_{\infty}}{(h_3)_{\infty}}.\]

Let $\lambda(x_{D_1})$ denote the sequence of $\lambda_i(x_{D_1})$ and
similarly $\lambda'(x_{D_1})$ denote the sequence of $\lambda'_i(x_{D_1})$, and  $\mu(x_{D})$ denote the sequence of $\mu_{ij} (x_{D_1})$.  Then $h_{\infty}$ can be determined from $\lambda'$ and $\mu$.

Then we can write
$S(c(h_1), s(a)) \prod_i S(c(b-\beta_i), s(a)) = A (\lambda (x_{D_1}), \mu (x_{D_1}), x_D)$ and
$ S(c(h_3), s(a))= B (\lambda' (x_{D_1}), x_D)$, and
since $S$ has degree $O(g)$ in each of the variable, it follows that
$A$ is polynomial in $x_D$ of degree $O(g^3)$ and rational in $x_{D_1}$  of degree $g^{O(g)}$, and
$B$ is polynomial in $x_D$ of degree $O(g^3)$ and rational in $x_{D_1}$ of degree $g^{O(g)}$.
Therefore
each coefficient of $h^+$ can be expressed as a rational function of degree $g^{O(g)}$ in $x_{D_1}$.

We have proved the following:

\begin{proposition}
\label{add-h}
Let $D_1=\Div (a_1,b_1)$ be a reduced divisor. Then $2D_1 = D' + (h)$ where $D'$ is a reduced divisor and $h$ is a function of the form ${h_1}{h_2}$ where $h_1 \in K[x]$ is of degree less than $2g$ and
$h_2 = \prod_i \frac{y - \beta_i (x)}{a'_i (x)} $ where $\beta_i (x)$ and $a'_i (x)$ are polynomials in $x$, and  $\deg \beta_i$ and the number of $i$ are both less than $2g$.  Put $h_3 (x)=\prod_i a'_i (x)$.  Then $d=\deg h_3 (x)=O(g^2)$.

\begin{enumerate}
\item
Let $h^+ (x_D)$ denote the function $h$ as it applies to evaluate the positive part $D^+$ of a reduced divisor $D=\Div (a,b)$.  Then

\[h^+ (x_D)= h(D^+) = \frac{S(c(h_1), s(a))}{S(c(h_3), s(a))}  {\prod_i S(c(b-\beta_i), s(a))}.\]

At each site of the doubling map, we have
\[
\begin{array}{lll}
h_1 (x) & = & \sum_{i=0}^{2g-1} \lambda_i (x_{D_1}) x^i\\
h_3 (x) & = & \sum_{i=0}^{d} \lambda'_i (x_{D_1}) x^i\\
\beta_i (x) & = & \sum_{j=0}^{2g-1} \mu_{ij}(x_{D_1}) x^j
\end{array}
\] where $\lambda_i$, $\lambda'_i$ and $\mu_{ij}$ are rational functions of degree $g^{O(g)}$ in $x_{D_1}$.

Let $\lambda(x_{D_1})$ denote the sequence of $\lambda_i(x_{D_1})$ and
similarly $\lambda'(x_{D_1})$ denote the sequence of $\lambda'_i(x_{D_1})$, and  $\mu(x_{D})$ denote the sequence of $\mu_{ij} (x_{D_1})$.  Then we can write
$S(c(h_1), s(a)) \prod_i S(c(b-\beta_i), s(a)) = A (\lambda (x_{D_1}), \mu (x_{D_1}), x_D)$ and
$ S(c(h_3), s(a))= B (\lambda' (x_{D_1}), x_D)$, where
$A$ is polynomial in $x_D$ of degree $O(g^3)$ and rational in $x_{D_1}$  of degree $g^{O(g)}$, and
$B$ is polynomial in $x_D$ of degree $O(g^3)$ and rational in $x_{D_1}$ of degree $g^{O(g)}$.

\item
$h_{\infty}
=\frac{\prod_{i,\deg\beta_i > g} (-\beta_i)_{\infty}}{(h_3)_{\infty}}$ and
$h_{\infty}$ can be determined from $\lambda' (x_{D_1})$ and $\mu (x_{D_1})$.
\end{enumerate}
\end{proposition}

We denote by $h_{D_1}$ the function $h$ in Proposition~\ref{add-h}, which is constructed by the addition algorithm.
\section{Pairing computation}
We keep the same notation as the last section and consider the pairing on $J[\ell]$ defined by Weil reciprocity.

If a reduced divisor $D$ represents an $\ell$-torsion point, then $\ell D$ is the divisor of a function $f$.  Given two reduced divisors $D_1$ and $D_2$ that represent two $\ell$-torsion points, we define the pairing to be
\[ e (D_1, D_2 ) = \frac{f_1 (D_2)}{f_2 (D_1)}\]
where $\ell D_i = (f_i)$ for $i=1,2$.

Let $D$ be a reduced divisor representing a point on $J$.  Then
$2D = (h_D) + D_1$ for some reduced divisor $D_1$.
For pairing computation we consider $h_D$ as a function that can evaluate at divisors of degree zero.
Thus on input a reduced divisor $D'=\Div (a',b')$, $h_D (D')\in\bar{k}$.
We note that any $a h_D$ with $a\in K^*$ defines the same function on divisors of degree zero.

Suppose $D$ is a $\ell$-torsion divisor.  We recall how to efficiently construct $h$ such that $\ell D = (h)$ through the squaring trick \cite{Miller,Miller1}.

Apply addition to double $D$, and get
\[ 2D = (h_D) + D_1 \] where $D_1$ is reduced.
Inductively, we have $H_i$ such that
\[ 2^i D = (H_i) + D_i \] with $D_i$ reduced.
Apply addition to double $D_i$ and get
\[ 2D_i = (h_{D_i}) + D_{i+1}\] with $D_{i+1}$ reduced.  Then
\[ 2^{i+1} D = (H_{i+1}) + D_{i+1}\]
where $H_{i+1} = H_i^2 h_{D_i}$.

Write $\ell = \sum_{i} a_i 2^i$ with $a_i \in \{0,1\}$.
There are $O(\log \ell)$ non-zero $a_i$.  So apply $O(\log\ell)$ many more additions and we can construct $h$ such that
$\ell D = (h)$.  Therefore if $D_1$ and $D_2$ are two reduced divisors, we can construct in this way
$f_1$ and $f_2$ such that $(f_i) = \ell D_i$ for $i=1,2$.  Moreover if we write $D_i=D_i^+ - r_i \infty$ for $i=1,2$,  then $\nu_{\infty} f_i = -\ell r_i$ for $i=1,2$.
So $\nu_{\infty} (f_1^{-r_2}f_2^{r_1})=0$.  Now it follows from Lemma~\ref{h-inf} that
\[
\frac{f_1(-r_2\infty)}{f_2(-r_1\infty)}=\alpha^{-r_2}\beta^{r_1}
\]
where $\alpha=(f_1)_{\infty}$ and $\beta=(f_2)_{\infty}$.

For $\hat{D}_1\in\hat{J} [\ell]$, $\hat{D}_2\in\hat{J}' [\ell]$
\[ \hat{e} (\hat{D}_1,\hat{D}_2) = \prod_{i=0}^{d-1} e_i ( \delta^{\sigma_i} \hat{D}_1,{\delta'}^{\sigma_i}\hat{D}_2)\]
where $e_i$ denotes the pairing defined by Weil reciprocity on $J^{\sigma_i}$.

\begin{lemma}
Let $\hat{D}\in\hat{J}(K)$. Suppose $2\hat{D}=\hat{D'}$ as points on $\hat{J}(K)$.  Then
\[ 2\delta\hat{D}^{\sigma_{-i}}=(h_{\delta\hat{D}^{\sigma_{-i}}})+\delta\hat{D'}^{\sigma_{-i}},\]
and
\[ 2\delta^{\sigma_i} \hat{D}=(h_{\delta\hat{D}^{\sigma_{-i}}}^{\sigma_i})+\delta^{\sigma_i}\hat{D'}.\]
\end{lemma}
\ \\{\bf Proof}
We have $\delta^{\sigma_i}\hat{D}=(\delta\hat{D}^{\sigma_{-i}})^{\sigma_i}$ with $\delta\hat{D}^{\sigma_{-i}}\in J(K)$.
Moreover since $2\hat{D} = \hat{D'}$ in $\hat{J}(K)$, we have $2\delta^{\sigma_i}\hat{D}=\delta^{\sigma_i} \hat{D'}$, and it follows that $2\delta\hat{D}^{\sigma_{-i}}=\delta\hat{D'}^{\sigma_{-i}}$, as points on $J$.  Therefore as reduced divisors
on $C$, we have
\[ 2\delta\hat{D}^{\sigma_{-i}}=(h_{\delta\hat{D}^{\sigma_{-i}}})+\delta\hat{D'}^{\sigma_{-i}}.\]
It follows that
\[ 2\delta^{\sigma_i} \hat{D}=(h_{\delta\hat{D}^{\sigma_{-i}}}^{\sigma_i})+\delta^{\sigma_i}\hat{D'}.\]
$\Box$

Suppose $2D = (h_D) + D_1$ as before.  By Proposition~\ref{add-h}, $h=h_D$ is of the form $h= \frac{h_1 (x) }{h_3 (x)} {\prod_i y-\beta_i (x)}$,
$h_{\infty} = \frac{\prod_{i, \deg \beta_i > g} -(\beta_i)_{\infty}}{(h_3)_{\infty}}$, and
\[
\begin{array}{lll}
h_1 (x) & = & \sum_{i=0}^{2g-1} \lambda_i (x_{D}) x^i\\
h_3 (x) & = & \sum_{i=0}^{d} \lambda'_i (x_{D}) x^i\\
\beta_i (x) & = & \sum_{j=0}^{2g-1} \mu_{ij}(x_{D}) x^j
\end{array}
\] where $\lambda_i$, $\lambda'_i$ and $\mu_{ij}$ are rational functions of degree $g^{O(g)}$ in $x_{D}$.

 Since $(h_3)_{\infty}$ and $(\beta_i)_{\infty}$ are determined respectively from their leading coefficients of the polynomials $h_3$ and $\beta_i$, hence they are determined by $\lambda'_i (x_D)$ and $\mu_{ij} (x_D)$.

Let
\[
\begin{array}{lll}
h'_1 (x) & = & \sum_{i=0}^{2g-1} \lambda_i (\delta\hat{D}^{\sigma_{-i}}) x^i\\
h'_3 (x) & = & \sum_{i=0}^{d} \lambda'_i (\delta\hat{D}^{\sigma_{-i}}) x^i\\
\beta'_i (x) & = & \sum_{j=0}^{2g-1} \mu_{ij}(\delta\hat{D}^{\sigma_{-i}}) x^j
\end{array}
\]

Then \[ h_{\delta\hat{D}^{\sigma_{-i}}} = \frac{h'_1 (x)}{h'_3 (x)}\prod_i (y-\beta'_i (x)).\]  So
\[ h^{\sigma_i}_{\delta\hat{D}^{\sigma_{-i}}} = \frac{{h'}_1^{\sigma_i} (x)}{{h'}_3^{\sigma_i} (x)} \prod_j (y-{\beta'}_j^{\sigma_i} (x)).\]

For $f=\lambda_i$, $\lambda'_i$ or $\mu_{ij}$, we have
\[( f (\delta\hat{D}^{\sigma_{-i}} )^{\sigma_i}  = f^{\sigma_i} (\delta^{\sigma_i} \hat{D}).
\]

So
\[
\begin{array}{lll}
{h'}^{\sigma_i}_1 (x) & = & \sum_{j=0}^{2g-1} \lambda^{\sigma_i}_j (\delta^{\sigma_i}\hat{D}) x^j\\
{h'}^{\sigma_i}_3 (x) & = & \sum_{j=0}^{d} {\lambda'}^{\sigma_i}_j (\delta^{\sigma_i}\hat{D}) x^j\\
\beta'_j (x) & = & \sum_{k=0}^{2g-1} \mu^{\sigma_i}_{jk}(\delta^{\sigma_i}\hat{D}) x^k
\end{array}
\]

It follows that $(h^{\sigma_i}_{\delta\hat{D}^{\sigma_{-i}}})_{\infty}$ can be determined from
$(\lambda'_j)^{\sigma_i} (\delta^{\sigma_i} \hat{D})$ and $(\mu_{jk})^{\sigma_i} (\delta^{\sigma_i} \hat{D})$,
which, as functions in $\hat{D}$.

We have proved the following.

\begin{lemma}
\label{h-delta-inf}
The function $(h^{\sigma_i}_{\delta\hat{D}^{\sigma_{-i}}})_{\infty}$ is determined by
$(\lambda'_j)^{\sigma_i} (\delta^{\sigma_i} \hat{D})$ and $(\mu_{jk})^{\sigma_i} (\delta^{\sigma_i} \hat{D})$.

\end{lemma}

Let $D=\Div (a,b)$ and $D'=\Div (a',b')$ be reduced divisors.
By Proposition~\ref{add-h}, we have
\[
h_D^+ (x_{D'}) = \frac{A(\lambda (x_D),\mu (x_D), x_{D'})}{B(\lambda' (x_D), x_{D'})}
\]
where $A (\lambda (x_{D}), \mu (x_{D}), x_{D'})=S(c(h_1), s(a')) \prod_i S(c(b'-\beta_i), s(a'))$ and
$B (\lambda' (x_{D}), x_{D'})= S(c(h_3), s(a'))$,
$A$ is polynomial in $x_D$ of degree $O(g^3)$ and rational in $x_{D_1}$  of degree $g^{O(g)}$, and
$B$ is polynomial in $x_D$ of degree $O(g^3)$ and rational in $x_{D_1}$ of degree $g^{O(g)}$.

Let $\hat{D}_1\in\hat{A}(K)$ and $\hat{D}_2\in\hat{A}' (K)$.
Then

\[h_{\delta\hat{D}_1^{\sigma_{-i}}} ( (\delta'\hat{D}_2^{ \sigma_{-i} })^+ ) =\frac{
 A( \lambda (\delta\hat{D}_1^{\sigma_{-i}} ),\mu (\delta\hat{D}_1^{\sigma_{-i}} ), \delta'\hat{D}_2^{ \sigma_{-i} }  )}
{B(\lambda' (\delta\hat{D}_1^{\sigma_{-i}} ), \delta'\hat{D}_2^{ \sigma_{-i} } )}
\]
It follows that
\[
h^{\sigma_i}_{\delta\hat{D}_1^{\sigma_{-i}}} ( ({\delta'}^{\sigma_i}\hat{D}_2)^+ )
 =
\frac{
 A^{\sigma_i}( \lambda^{\sigma_i} (\delta^{\sigma_i} \hat{D}_1  ), \mu^{\sigma_i}_1 (\delta^{\sigma_i} \hat{D}_1  ), \delta'^{\sigma_i}\hat{D}_2  )}
{B^{\sigma_i}(\lambda'^{\sigma_i} ( \delta^{\sigma_i}\hat{D}_1 ), \delta'^{\sigma_i}\hat{D}_2 )}
\]

Write $A/B$ in the form
\[ \frac{A(\lambda (x_D),\mu (x_D), x_{D'})}{B(\lambda' (x_D), x_{D'})}=\frac{A_1(x_D, x_{D'})}{B_1(x_D, x_D')}\]
where $A_1$ and $B_1$ are polynomials in $x_D$ and $x_{D'}$.

Let $\varphi: J\times J \to \bar{k}$ be the rational function defined by $A_1/B_1$.  Then

We have
\[
h^{\sigma_i}_{\delta\hat{D}_1^{\sigma_{-i}}} ( ({\delta'}^{\sigma_i}\hat{D}_2)^+ )
 =
\frac{A_1^{\sigma_i}( \delta^{\sigma_i} \hat{D}_1  , \delta'^{\sigma_i}\hat{D}_2  ) }{B_1^{\sigma_i}( \delta^{\sigma_i} \hat{D}_1  , \delta'^{\sigma_i}\hat{D}_2  )}=\varphi^{\sigma_i} (\delta^{\sigma_i}\hat{D}_1,{\delta'}^{\sigma_i}\hat{D}_2)
\]

We have proved the following:

\begin{lemma}
\label{h-delta}
Let $A$ and $B$ be polynomials as defined in Proposition~\ref{add-h}.  Write $A/B$ in the form
\[ \frac{A(\lambda (x_D),\mu (x_D), x_{D'})}{B(\lambda' (x_D), x_{D'})}=\frac{A_1(x_D, x_{D'})}{B_1(x_D, x_D')}\]
where $A_1$ and $B_1$ are polynomials of degree $O(g^3)$ in $x_D$ and degree $g^{O(g)}$ in $x_{D'}$.
Let $\varphi: J\times J \to \bar{k}$ be the rational function defined by $A_1/B_1$.  Then
$
\varphi^{\sigma_i} (\delta^{\sigma_i}\hat{D}_1,{\delta'}^{\sigma_i}\hat{D}_2)=h^{\sigma_i}_{\delta\hat{D}_1^{\sigma_{-i}}} ( ({\delta'}^{\sigma_i}\hat{D}_2)^+ )$.
\end{lemma}

Put $\hat{D}_0 = \hat{D}$.  Suppose inductively
$2\hat{D}_j = \hat{D}_{j+1}$.
Inductively, we have $H^{(i)}_j$ such that
\[ 2^j \delta^{\sigma_i}\hat{D} = (H^{(i)}_j) + \delta^{\sigma_i}\hat{D}_j \] with $\hat{D}_j$ reduced.
Apply addition to double $\delta^{\sigma_i}\hat{D}_j$ and get
\[ 2\delta^{\sigma_i}\hat{D}_j = (h^{\sigma_i}_{\delta \hat{D}_j^{\sigma_{-i}}}) + \delta^{\sigma_i}\hat{D}_{j+1}\] with $\delta \hat{D}_{j+1}$ reduced.  Then
\[ 2^{j+1} \delta^{\sigma_i}\hat{D} = (H^{(i)}_{j+1}) + \delta^{\sigma_i}\hat{D}_{j+1} \]
where $H^{(i)}_{j+1} = (H^{(i)}_j)^2 h^{\sigma_i}_{\delta\hat{D}_j^{\sigma_{-i}}}$.

Write $\ell = \sum_{i} a_i 2^i$ with $a_i \in \{0,1\}$.
There are $O(\log \ell)$ non-zero $a_i$.  So apply $O(\log\ell)$ many more additions and we can construct $H_{\hat{D}}^{(i)}$ such that
$\ell \delta^{\sigma_i}\hat{D}  = (H_{\hat{D}}^{(i)})$.

For $\hat{D}_1,\hat{D_2}\in\hat{A}[\ell]$,
\[ e_i ( \delta^{\sigma_i}\hat{D_1}, {\delta'}^{\sigma_i}\hat{D_2} ) = \prod_i \left( \frac{H_{\hat{D_1}}^{(i)}({\delta'}^{\sigma_i}\hat{D_2})}{H_{\hat{D_2}}^{(i)}(\delta^{\sigma_i}\hat{D_1})} \right)^{a_i}.\]

In summary, to specify the program for $\hat{e}$, it is enough to specify $O(g^2 d)$ many descent functions with blinded constant factors of a set $C$ of $O(g^2)$ rational functions on $J$.  The set $C$ contains the following functions
\begin{enumerate}
\item $O(g)$ functions that define $m: J\times J\to J$,
\item $\varphi : J\times J \to \bar{k}$ such that $\varphi (x_D, x_{D'}) = h_D ({D'}^+)$,
\item $O(g^2)$ functions ( ${\lambda'}_i$, $\mu_{ij}$) from which $(h_D)_{\infty}$ can be determined.
\end{enumerate}
The following descent functions are specified:
$\varphi\circ (\delta,{\delta'})$,  $(\lambda'_j) \circ\delta $ and $\mu_{jk}\circ \delta$
where $h^{\sigma_i}_{\delta\hat{D}^{\sigma_{-i}}} ( ({\delta'}^{\sigma_i}\hat{D'})^+ )=\varphi^{\sigma_i}(\delta^{\sigma_i}\hat{D},{\delta'}^{\sigma_i}\hat{D'} )$, $\varphi (x_D,x_{D'}) =\frac{A_1 (x_D, x_{D'})}{B_1(x_D,x_{D'})}$ and $A_1$ and $B_1$ are polynomials of degree $O(g^3)$ in $x_D$ and degree $g^{O(g)}$ in $x_{D'}$.

The associated descent functions with blinded constant factors can all be specified properly.

For $g >1$, the degrees of the functions $f\in C$ are all greater than 1.  Perform a birational transformation $\iota$ as described in \S~\ref{birational-model} if necessary, we can replace these functions $f\in C$ by functions $f\circ \iota^{-1}$ which are likely dense for some degree at least 2, hence are immune to the linear attack described in \S~\ref{linear-analysis}.  Moreover $f\circ\iota^{-1}$ is secret since $\iota$ is secretly chosen.

We have proved the following:

\begin{theorem}
Efficient computation for the blinded pairing $\hat{e}_{\ell}$ on $\hat{J}[\ell]$ can be properly specified, such that the specification does not contain any global descent.  More precisely, to specify the program for $\hat{e}$, it is enough to specify $O(g^2 d)$ many descent functions with blinded constant factors of a set $C$ of $O(g^2)$ rational functions on $J$.  For $g >1$, the degrees of the functions $f\in C$ are all greater than 1.  Suppose by performing a birational transformation $\iota$ as described in \S~\ref{birational-model} if necessary, and the functions $f\circ \iota^{-1}$, which replaces $f\in C$, are dense for some degree at least 2.  Then the specification is safe from the linear attack described in \S~\ref{linear-analysis}.  Moreover $f\circ\iota^{-1}$ for $f\in C$ is secret since $\iota$ is secretly chosen.
\end{theorem}

\section{The elliptic curve case}
\label{elliptic}
In this section we specialize the trilinear map construction to the case where the dimension of the abelian variety $A$ is one, namely the elliptic curve case.
We take the abelian variety to be an elliptic curve $E$ defined over $K$.
Suppose the characteristic of $K$ is not 2 or 3, and $E$ is given $y^2 =x^3 +ax+b$ with $a,b\in K$.
A reduced divisor $(x-a,b)$ in this case corresponds to an a point $(a,b)\in E(\bar{k})$, and the reduced divisor
$(1,0)$ corresponds to the zero point of $E$ ({\em the point at infinity}), which is not on the affine model
$y^2 = x^3+ax+b$.  In this case the correspondence between a reduced divisor and a point is very direct.
We can regard $E(\bar{k})$ as consisting of an affine piece $V=\{(1,x,y): y^2 = x^3 +ax+b\}$, and a zero point $(0,1,0)$.
When we deal with nonzero points we can simply identify $V$ with the curve $y^2 = x^3 +ax+b$, which is the principal site of $E$.

The addition map can be described as follows (see \cite{Sil}).
Let $P_1 = (x_1,y_1)$, $P_2 = (x_2,y_2)$ be two points on $V$.
If $x_1 = x_2$ and $y_1 = -y_2$, then $P_1 + P_2 = 0$.
Otherwise, we can find $P_3 = (x_3,-y_3)$ such that
$P_1$, $P_2$ and $\bar{P}_3 = (x_3,y_3)$ lie on a line
$y = \lambda x + \nu$, and we have
$P_1 + P_2 = P_3$.

(1) If $x_1 \neq x_2$, then $\lambda = \frac{y_2-y_1}{x_2-x_1}$ and $\nu = \frac{y_1 x_2 - y_2 x_1}{x_2-x_1}$.

(2) If $x_1 =  x_2$ and $y_1 \neq 0$, then $\lambda = \frac{3x_1^2 +a}{2y_1}$ and $\nu = \frac{- x^3_1+ax_1+2b}{2y_1}$

In both cases
$x_3 = \lambda^2 -x_1-x_2$, $y_3 = -\lambda x_3 -\nu$.

The principal site of the addition map corresponds to Case (1) and consists of $(P_1,P_2)\in V\times V$ where $P_1 = (x_1,y_1)$, $P_2 = (x_2,y_2)$ and $x_1\neq x_2$, corresponding to Case (1).

The principal site of the doubling map corresponds to Case (2) and consists of $P=(x_1,y_1) \in V$ where $y_1\neq 0$.

Note that for doubling map all but the two torsion points are at the principal site.
For $D=P_1 - \infty$ where $P_1 = (x_1,y_1)$ is not 2-torsion, we have $2D = (h_D) + D'$ where $D' = P_3-\infty$ with
$P_3 = (x_3,y_3)$ given by the formula above, and $h_D (x.y)  = \frac{L}{x-x_1}$ where $L=y-\lambda x -\nu$,
$\lambda = \frac{3x_1^2 +a}{2y_1}$ and $\nu = \frac{- x^3_1+ax_1+2b}{2y_1}$.

Observe that in this situation $(h_D)_{\infty}=1$, therefore for pairing computation we only need to focus on $h_D$ as a function on the affine points, namely $(x,y)\in V$.

The degree of the addition map is of degree at least 2 in $x_1,y_1$ and in $x_2,y_2$.  The degree of the doubling map is of degree at least 2 in $x$ and $y$ as well. However  the degree of $h_D$ in $x$ (resp. $y$) is 1.  This raises the concern of the possibility of linear attack discussed in \S~\ref{linear-analysis}.   To prevent such an attack we consider birational models of $y^2=x^3+ax+b$ where the corresponding function for $h_D$ is of degree greater than 1 in all variables.  We consider one such model
$$E_1: \left( \frac{\alpha x_1^2}{\beta x_1 + \gamma x_2}\right)^2 = x_1^3 + ax_1 + b$$
with the birational map  $\iota_1:E_1 \to E$ sending $(x_1,x_2)\in E_1$ to
$(x,y)\in E$ where $x= x_1$,
$y=\frac{\alpha x_1^2}{\beta x_1 + \gamma x_2}$ and $\alpha,\beta\in K^*$.
Let $(x_1,x_2), (u_1, u_2)\in E_1$, and $D_1 = (u_1,u_2)-\infty$.

For $E$ we have $h_D (x,y)= \frac{y-\lambda x -\nu}{x-u}$, where $D=(u,v)-\infty$, $\lambda = \frac{3u^2 +a}{2v}$ and $\nu = \frac{- u^3+au+2b}{2v}$.   For $E_1$, $h_D$ is replaced by
\[ h^{E_1}_{D_1} (x_1,x_2):=h_{\iota_1 D_1} (\iota_1 (x_1,x_2)) = \frac{ \frac{\alpha x_1^2}{\beta x_1 + \gamma x_2}-\lambda' x_1 -\nu'}{x_1-u_1}\]
where
\[ \lambda'=\lambda(\iota_1 (u_1,u_2))= \frac{3u_1^2 +a}{2\frac{\alpha u_1^2}{\beta u_1 + \gamma u_2}}
= \frac{(3u_1^2 +a)(\beta u_1+\gamma u_2)}{2\alpha u_1^2}\]

\[ \nu'=\nu(\iota_1 (u_1,u_2))=
= \frac{(-u_1^3 +au_1+2b)(\beta u_1+\gamma u_2)}{2\alpha u_1^2}\]

We see that $h^{E_1}_{D_1} (x_1,x_2)$ is of the form $A(u_1,u_2,x_1,x_2)/B(u_1,u_2,x_1,x_2)$ where $A$ is of degree $2$ in $x_1$ and $x_2$ and degree 4 in $u_1$ and $u_2$, and $B$ is of degree $2$ in $x_1$ and $x_2$ and degree 3 in $u_1$ and $u_2$.

Let $m:E\times E \to E$ be the addition map as described above for the principal site.  Then the addition map $m_1:E_1\times E_1\to E_1$ for $E_1$ is $i_1^{-1}\circ m\circ (i_1,i_1)$.  For $(x_1,y_1, x_2,y_2)$ at the principal site where $(x_1,y_1)\in E_1$ and $(x_2,y_2)\in E$, write $m_1 (x_1,y_1,x_2,y_2)= (x_3,y_3)$, then the formula for $x_3$ and $y_3$ as rational functions in $x_1,y_1,x_2,y_2$ can be similarly worked out.  The function describing $x_3$ has the form $F/G$ where $F(x_1,y_1,x_2,y_2)$ is of total degree 7, of degree 5 in $x_1,y_1$, of degree 5 in $x_2,y_2$, and
$G(x_1,y_1,x_2,y_2)$ is of total degree 6, of degree 4 in $x_1,y_1$, and of degree 4 in $x_2,y_2$.
The function describing $y_3$ has the form $F/G$ where $F(x_1,y_1,x_2,y_2)$ is of total degree 16, of degree 11 in $x_1,y_1$, of degree 11 in $x_2,y_2$, and
$G(x_1,y_1,x_2,y_2)$ is of total degree 15, of degree 10 in $x_1,y_1$, of degree 10 in $x_2,y_2$.

Similarly Let $\tau:E \to E$ be the doubling map as described above for the principal site.  Then the doubling map $\tau_1:E_1\to E_1$ for $E_1$ is $i_1^{-1}\circ \tau\circ (i_1,i_1)$. For $(x,y)$ at the principal site of $\tau$  where $(x,y)\in E_1$, write $\tau_1: (x,y)= (x',y')$, then the formula for $x'$ and $y'$ as rational functions in $x,y$ can be similarly worked out.  The function describing $x'$ has the form $F/G$ where $F(x,y)$ is of total degree 6,  and
$G(x,y)$ is of total degree 2. The function describing $y'$ has the form $F/G$ where $F(x,y)$ is of total degree 15,  and
$G(x,y)$ is of total degree 11.

Perform a random birational transformation $\iota_2:E_1\to E_2$ as described in \S~\ref{birational-model} and suppose
$\iota_2^{-1} (z) = ( L_1 (z), L_2 (z) )$  for $z=(z_1,z_2,z_3)$, where $L_1$ and $L_2$ are randomly chosen linear forms over $K$.  Suppose $z,z'\in E_2$, and let $D_z = z-\infty$, where $z=(z_1,z_2,z_3)$ and $z'=(z'_1,z'_2,z'_3)$.
Then $h^{E_2}_{D_z} (z')$ is of the form $\varphi (z,z')=A'(z,z')/B'(z,z')$ where $A'(z,z')=A(L_1(z),L_2 (z),L_1 (z'),L_2 (z'))$
and $B'(z,z')=B(L_1(z),L_2 (z),L_1 (z'),L_2 (z'))$.
Both $A'$ and $B'$ are likely dense for some degree at least 2 in $z$ and in $z'$, in which case they are both safe for specification.

The addition map for $E_2$ is $i_2^{-1}\circ m_1\circ (i_2,i_2)$.   Similarly
the doubling map for $E_2$ is $i_2^{-1}\circ \tau_1\circ (i_2,i_2)$.  The map $m_2$ (resp. $\tau_2$) is defined by 3 rational functions.  As they are formed with randomly chosen $i_2$, the polynomials describing them are likely dense for some degree at least 2, hence safe for specification.

Therefore, to specify the program for $\hat{e}$ on $E_2$, it is enough to specify $O(d)$ many descent functions with blinded constant factors of a set of 7 rational functions, three functions that define the addition map on $E_2$, three functions that define the doubling map, and one function $\varphi (z,z')$ for defining  $h^{E_2}_{D_z} (z')$. These 7 functions are secret since $\iota_1$ and $\iota_2$ are secret. The $O(d)$ descent functions and maps can  be specified properly such that the specification contains no global descent.

We have proved the following:

\begin{theorem}
Efficient computation for the blinded pairing $\hat{e}_{\ell}$ on $\hat{E}[\ell]$ can be properly specified, such that the specification does not contain any global descent.  More precisely, to specify the program for $\hat{e}$, it is enough to specify $O(d)$ many descent functions with blinded constant factors of a set of 7 rational functions.  By performing a random secret birational transformation the 7 functions are secret and likely dense for some degree at least 2, in which case the specification is safe from the linear attack described in \S~\ref{linear-analysis}.
\end{theorem}

\section{Open problems}
\label{open}
We summarize several computational problems which are important to the security of the trilinear map discussed in this paper. For simplicity we focus on the elliptic curve case and assume only one secret descent basis is used in the construction instead of two.  The discussion naturally extends to the general case of Jacobian varieties of hyperelliptic curves and where two descent bases are used.

Let $\bu$ be a randomly chosen basis of $K$ over $k$.
Suppose $E$ is an elliptic curve defined over $K$ and $\mu_{\ell}\subset K$.  Suppose we have formed a birational model $E_2$ of $E$ as in \S~\ref{elliptic} such that the two polynomials defining $E_2$ are dense in degree 2.  Moreover the rational function $\varphi$ describing  $h^{E_2}_{D_z} (z')$ for computing $e_{\ell}$, and the 6 functions describing the addition map and the doubling map are safe for specification.

We now call $E_2$ as $E$ and $m_2$ as $m$, and $\tau_2$ as $\tau$.

As in \S~\ref{trapdoor-dl}, we choose a set of $N=O(d^2)$ many $(0,1)$-matrices $M_1,\ldots,M_N$ that span $Mat_d (\F_{\ell})$, so that each $M=M_i$ has the following properties:
\begin{enumerate}
\item
There are exactly two nonzero entries $(i,i_1)$ and $(i,i_2)$ for row $i$, for $i=0,\ldots,d-1$.
\item
Let $E_M=\{ (i-i_1\mod d, i-i_2\mod d : i=0,\ldots,d-1\}$.  For $(a,b)\in E_M$, let
$I_{M,a,b}=\{i: (i-i_1,i-i_2) = (a,b)\mod d \}$.
Then $| I_{M_i,a,b} | = \Theta(d^{\epsilon}) $ for all $(a,b)\in E_{M_i}$, for some positive constant $\epsilon < 1$,
\end{enumerate}
Again,
we have the following commutative diagram:

\[
\begin{array}{llll}
\hat{E}[\ell] & \stackrel{\rho}{\to} & \prod_{i} E^{\sigma_i} [\ell]\stackrel{\prod_i \sigma_{-i}}{\to}  & E[\ell]^d\\
\downarrow \Psi_i & &   & \downarrow \varphi_{M_i}\\
\hat{E}[\ell] & \stackrel{\rho}{\to} & \prod_{i} E^{\sigma_i} [\ell] \stackrel{\prod_i \sigma_{-i}}{\to} & E[\ell]^d

\end{array}
\]

Let $C=\{ (i-i_1\mod d, i-i_2\mod d : i=0,\ldots,d-1\}$.  For $(a,b)\in C$, let
$I_{a,b}=\{i: (i-i_1,i-i_2) = (a,b)\mod d \}$.
Let $\Omega_{a,b}=W^{I_{a,b}} \Gamma_{I_{a,b}}$.
By Proposition~\ref{specify-psi}, the set of $\Psi_i$ can be specified
 by specifying $\hat{m}$, making public $C$, and $\Omega_{a,b}$ for every $(a,b)\in C$.

 From the published information: $D_{\alpha},D_{\beta}\in \hat{E}[\ell]$, specified $\hat{m}$, specified $\hat{\tau}$,
 specified $\varphi\circ\delta$, and the set of $\Omega_{a,b}$, $(a,b)\in C$,
 can $\bu$ be efficiently determined?

From the above-mentioned published information, together with the set $\cR$ of quadratic relations on the matrices $M_i$ described in \S~\ref{trapdoor-dl}, can the trapdoor discrete-log problem on $G_3$ be solved efficiently?

The following problems that do not involve the pairing computation can be separated out and more narrowly defined.  Solving any one of these problems efficiently will break the trilinear map.
\begin{enumerate}
\item
Assume the birational model $E$ for an elliptic curve, the addition map $m:E\times E\to E$ and the doubling map $\tau :E\to E$ are safe for specification, and $\hat{m}$ and $\hat{\tau}$ are properly specified (for the principal sites, each by $3d$ polynomials of degree $O(1)$ in $3d$ variables over $k$).  Can $\bu$ be determined efficiently?
\item
A set $S$ of $O(d^2)$ subsets of $\{0,\ldots,d-1\}$ is secretly chosen, each subset $I$ is of cardinality $\Theta(d^{\epsilon}) $ for some positive constant $\epsilon < 1$.  The set of matrices $W^{I} \Gamma_{I}$ is made public ($\Gamma$ is the matrix whose $i$-th row is $\bu^{\sigma_i}$ for $i=0,\ldots,d-1$, and $W=\Gamma^{-1}$).  Can
$\bu$ be determined efficiently?  This problem is abstracted as a subproblem from the next problem.
\item
The trapdoor discrete logarithm problem as described in \S~\ref{trapdoor-dl}, both the generic version, which does not involve $A=E$, and the version that allows public identity testing, which involves $A=E$ but does not involve pairing. We remark that the generic version can be reduced to solving a system of quadratic polynomials in $d^{O(1)}$ variables. However the best known method for solving such systems has time complexity exponential in the number of variables.
\end{enumerate}

Whether or a not there is a secure trilinear map without a trapdoor is an interesting open problem.  The approach in \cite{H},which proceeds more closely along the line suggested by Chinburg and does not involve Weil descent, remains to be further investigated.

\section*{Acknowledgements}
I would like to thank the participants of  the AIM workshop on cryptographic multilinear maps (2017), and the participants of the BIRS workshop: An algebraic approach to multilinear maps for cryptography (May 2018), for stimulating and helpful discussions.

I would especially like to acknowledge the contributions of the following colleagues:
Dan Boneh and Amit Sahai for valuable discussions during the early phase of this work;
Steven Galbraith for careful reading of the preprint in \cite{H} as well as valuable comments and questions;
Steven Galbraith and Ben Smith for valuable comments and questions on a subsequent preprint \cite{H1};
Karl Rubin, Shahed Sharif, Alice Silverberg, and Travis Scholl for reading \cite{H1} and for many helpful discussions leading up to the current revision.

\end{document}